\documentclass[12pt, draftclsnofoot, onecolumn]{IEEEtran}
\usepackage{soul}            
\usepackage{multicol}
\usepackage{epsfig}
\usepackage{epstopdf}
\usepackage{float}
\usepackage{subfig}
\usepackage{perpage}
\MakeSorted{table}
\usepackage{array}
\usepackage{relsize}
\usepackage{tikz}
\usepackage[linesnumbered,ruled]{algorithm2e}
\usepackage{enumitem}
\usepackage{graphicx}
\usepackage{amsfonts}
\usepackage{amssymb}

\usepackage[T1]{fontenc}
\usepackage{etoolbox}
\usepackage[noend]{algpseudocode}
\usepackage{cite}
\DeclareMathAlphabet\mathbfcal{OMS}{cmsy}{b}{n}
\usepackage[cmex10]{amsmath}
\usepackage{array}
\usepackage{fixltx2e}
\usepackage{tabularx}
\usepackage{wrapfig}
\usepackage[hidelinks]{hyperref}
\usepackage{mathtools, cuted}
\usepackage{hhline}

    \newcolumntype{L}{>{\raggedright\arraybackslash}X}

\newcommand{\argmax}{\arg\!\max}
\usepackage{array}

\newtheorem{lemma}{Lemma}
\ifCLASSINFOpdf
\else
\fi

\usepackage{etoolbox} 
\newtoggle{SINGLE_COL}
\toggletrue{SINGLE_COL}     
\togglefalse{SINGLE_COL}   

\makeatletter
\def\BState{\State\hskip-\ALG@thistlm}
\makeatother

\newtheorem{remark}{Remark}
\begin{document}
\bstctlcite{IEEEexample:BSPcontrol}
\title{OTFS  Channel Estimation And Data Detection Designs With Superimposed Pilots}

\author{\IEEEauthorblockN{Himanshu B. Mishra, Prem Singh, Abhishek K. Prasad, and Rohit Budhiraja
\thanks{Himanshu B. Mishra and Abhishek K. Prasad are with the Department of Electronics Engineering, IIT Dhanbad, India (e-mail: himanshu$\emph{@}$iitism.ac.in, akp.1462$\emph{@}$gmail.com).Prem Singh and Rohit Budhiraja are with the Department of Electrical  Engineering, IIT Kanpur, 208016, India (e-mail: \{psrawat, rohitbr\}$\emph{@}$iitk.ac.in).}}
}

\IEEEtitleabstractindextext{
\vspace{-1.8cm}
\begin{abstract}
This work proposes a superimposed pilot (SP)-based channel estimation and data detection framework for orthogonal time-frequency space (OTFS) scheme, wherein low-powered pilots are superimposed on to data symbols in the delay-Doppler domain.  We propose two channel estimation and data detection designs for SP-OTFS systems which,  unlike the existing OTFS designs, do not designate any slots for pilots, which improves their spectral efficiency (SE). The first SP design estimates channel  by treating data as interference, which degrades its performance at high signal to noise ratio. The second SP design alleviates this problem by iterating between channel estimation and  data detection. Both these designs detect data using message passing algorithm which exploits OTFS channel sparsity, and consequently has low computational complexity. We also derive a lower bound on the signal-to-interference-plus-noise ratio of the proposed designs, and maximize it  by optimally allocating power between data and pilot symbols.  We numerically validate the derived analytical results, and show that the proposed designs have superior SE than the two  state-of-the-art OTFS channel estimation and data detection designs.
\end{abstract}

\vspace{-0.4cm}
\begin{IEEEkeywords}
\vspace{-0.4cm}
Message passing algorithm, orthogonal time-frequency space (OTFS) scheme, superimposed pilots.
\end{IEEEkeywords}}

\maketitle

\IEEEdisplaynontitleabstractindextext

\IEEEpeerreviewmaketitle
\vspace{-1.2cm}
\section{Introduction}
A practical wireless channel, due to multi-path propagation  and Doppler shift, is both time- and frequency-selective. The cyclic prefix (CP)-aided orthogonal frequency division multiplexing (OFDM)  combats the  channel frequency-selectivity \cite{DBLP:journals/twc/WangPMZ06}. A large Doppler shift, due to high-speed relative movement between the transmitter and receiver, disturbs the inter-subcarrier orthogonality  in an OFDM system, which significantly degrades its performance \cite{DBLP:journals/twc/WangPMZ06}. The orthogonal time frequency and space (OTFS) scheme is designed to improve performance  in high-Doppler scenarios by multiplexing transmit  symbols in the delay-Doppler domain \cite{DBLP:journals/twc/SurabhiAC19, SurabhiC20,raviteja2018interference, DBLP:conf/ita/MuraliC18, DBLP:journals/tit/FishGHSS13,DBLP:conf/globecom/RamachandranC18}. 
This is unlike OFDM which multiplexes them in the time-frequency domain. The OTFS scheme, by using inverse symplectic finite Fourier transform (ISFFT) and Heisenberg transform at the transmitter, and their inverse at the receiver, converts a  doubly-selective channel into an almost time-invariant  one, in the delay-Doppler domain. OTFS symbols, thus, experience almost-constant channel gain, which can be exploited to reduce  pilot overhead  for estimating rapidly time-varying channel.  Further, the delay-Doppler domain channel, due to small number of clusters, is sparse. This can also be exploited to reduce channel estimation and data detection complexity \cite{raviteja2018interference,DBLP:conf/ita/MuraliC18, DBLP:journals/tit/FishGHSS13,DBLP:conf/globecom/RamachandranC18,DBLP:journals/tvt/RavitejaPH19}.





G. D. Surabhi \textit{et al.} in \cite{DBLP:journals/twc/SurabhiAC19} derived the diversity of single-input single-output (SISO)/multiple-input multiple-output (MIMO)-OTFS systems using maximum-likelihood (ML) decoding with perfect receive channel state information (CSI). 
	The authors in \cite{SurabhiC20} developed linear equalizers for SISO OTFS systems.
 Raviteja \textit{et al.} in \cite{raviteja2018interference} proposed a computationally-efficient detector based on message passing  algorithm, which exploits OTFS channel sparsity. These works assumed perfect receive CSI which, however, needs to be estimated in  practice. {The perfect CSI assumption also simplifies their message calculations.}
{References \cite{DBLP:conf/ita/MuraliC18, DBLP:journals/tit/FishGHSS13} designed pilot-based OTFS  channel estimators in  time-frequency domain which, due to  non-sparse time-frequency channel, are computationally complex. Also, the time-frequency channel estimate needs to be transformed into the delay-Doppler domain for incorporating low-complexity message passing receiver.} The authors in \cite{DBLP:conf/globecom/RamachandranC18, DBLP:journals/tvt/RavitejaPH19} developed channel estimators in the delay-Doppler domain. In particular, Ramachandran \textit{et al.} in \cite{DBLP:conf/globecom/RamachandranC18} first estimated the OTFS channel using a frame  consisting entirely of pilot symbols, and used it in the subsequent frames to detect data. This degrades the system spectral efficiency (SE), as an entire OTFS frame is used for estimating channel. Furthermore, its data detection performance may  degrade  due to extremely high Doppler spread, which leads to channel aging between two frames. 

\begin{figure}[htbp]
	\begin{center}
		\subfloat[]{\includegraphics[scale = 0.6]{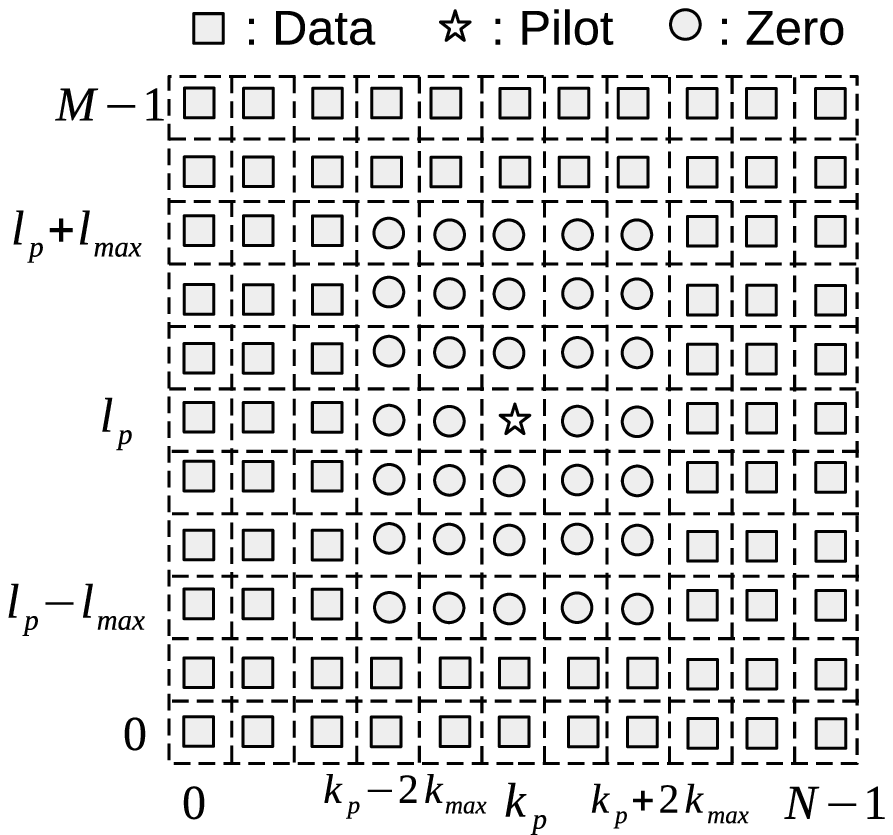}}
		\hfil 
		\hspace{-30pt}\subfloat[]{\includegraphics[scale = 0.6]{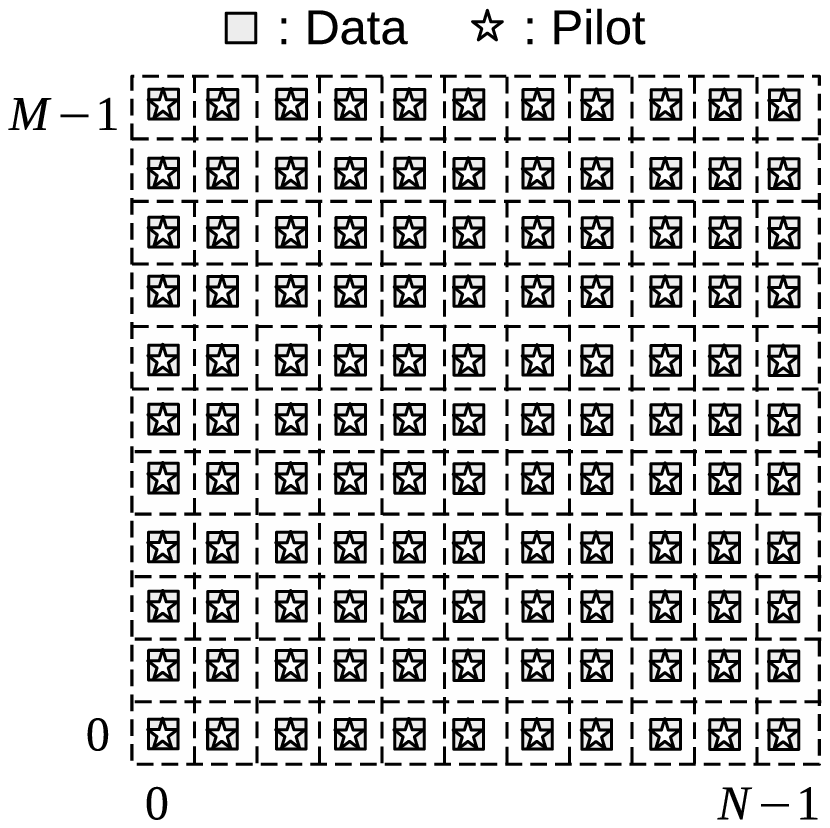}}
	\end{center}	
	\hfil
	\vspace{-0.9cm}
	\caption{\small Frame structure in OTFS systems: (a) EP-based design in \cite{DBLP:journals/tvt/RavitejaPH19}; and (b) proposed SP-aided designs.}
	\label{fig:Frame}
		\vspace{-1cm}
\end{figure}

Raviteja \textit{et al.} in \cite{DBLP:journals/tvt/RavitejaPH19} proposed an embedded pilot (EP)-based OTFS channel estimator for the frame structure  which, as shown in Fig.~\ref{fig:Frame}(a), consists of both pilot and data. 
Here $(l_p, k_p)$ gives the pilot location at the $l_p$th delay, and the $k_p$th Doppler tap. This design first estimates the channel using the pilot symbols,  and subsequently detects  data using the estimated channel. This frame structure necessitates the insertion of zeros between the pilots and data symbols to avoid mutual interference between them \cite{DBLP:journals/tvt/RavitejaPH19}. The number of inserted zeros depends on the values of $l_{max}$ and $k_{max}$, which are the delay and Doppler taps corresponding to their maximum value. The insertion of zeros reduces the SE, especially when the maximum delay value and the Doppler shift are high, a fact we will numerically verify in the sequel.

This SE loss, due to conventional pilot-based channel estimation \cite{DBLP:conf/globecom/RamachandranC18,DBLP:journals/tvt/RavitejaPH19}, can be reduced by superimposing pilots on to  information symbols. {Reference \cite{Carrasco-AlvarezPOT12} proposed a superimposed training (ST)-aided method for estimating doubly-selective channels in  time-frequency domain for a single carrier system. The estimator therein projects  a subspace of the time-varying channel onto a set of two dimensional orthogonal functions.} References \cite{tugnait2003channel, he2007superimposed, DBLP:journals/tsp/HeT08} used idea of ST for single-/multi-antenna channel estimation in time-frequency domain. {These works \textit{crucially} considered a nearly time-invariant channel that is constant over a large number of frames, and used  a periodic superimposed pilot (SP) with zero-mean data symbols. This enabled them to mitigate the mutual interference between data and pilot symbols using the first-order statistics of the received signal. The authors in \cite{TranPTN09,TranTN10} proposed affine-precoder-based SP channel estimator, which, unlike \cite{tugnait2003channel, he2007superimposed, DBLP:journals/tsp/HeT08}, completely removes the interference between data and pilots at the cost of reduced SE.  The designs in \cite{tugnait2003channel, he2007superimposed, DBLP:journals/tsp/HeT08,TranPTN09,TranTN10}
also assumed \textit{time-invariant channel}, which degrades their performance for rapidly time-varying channels.

The data-dependent superimposed training (DDST) scheme superimposes, in addition to pilots,  the arithmetic mean of transmit data on the transmitted information symbols \cite{GhoghoMAS05}. The DDST scheme, as explained in \cite{Longoria-Gandara16}, has data identifiability problem.   Reference \cite{Longoria-Gandara16} resolved it by proposing a hybrid solution, known as joint mean removal SP and pilot-aided training (PAT) for detection and channel estimation in MIMO system over flat-fading quasi-static channels. The design in \cite{Longoria-Gandara16}, similar to \cite{tugnait2003channel, he2007superimposed, DBLP:journals/tsp/HeT08}, requires channel to be time-invariant for a large number of frames, to compute the time average of received signals. The SP-aided transmission in OFDM cannot be trivially extended for the delay-Doppler channel estimation in OTFS systems as the i) channel changes rapidly in the time-frequency domain; and ii) channel gain in the delay-Doppler domain varies across frames \cite{DBLP:journals/tvt/RavitejaPH19}.  An OTFS system, unlike its OFDM counterpart processes the complete transmit frame of size $M\times N$ where $M$ and $N$, , as shown in Fig.~\ref{fig:Frame},  denote the number of delay and Doppler bins \cite{raviteja2018interference}. This considerably increases the OTFS system complexity as its channel matrix is of $MN\times MN$ size. \textit{Both channel estimator and data detector in OTFS  need to be designed by exploiting delay-Doppler channel sparsity which will reduce their complexities.} 

To extend SP framework to the OTFS systems, it is therefore necessary to estimate channel and detect data within a frame, and that too in  presence of interference between data and pilot symbols in the delay-Doppler domain \cite{DBLP:journals/tvt/RavitejaPH19}. {Also, the conventional data detection techniques do not exploit the OTFS sparsity, and are therefore not computationally-efficient \cite{raviteja2018interference}.  Further, to reduce the bit error rate (BER) and to increase the  SE, the transmit power need to be optimally allocated between the data and pilot symbols.  Given the aforementioned key challenges, the aim of this work is to develop SP-aided channel estimation and data detection designs in the delay-Doppler domain for OTFS systems}. To the best of our knowledge, SP-aided OTFS systems have not been investigated in the existing literature. The \textbf{main contributions} of this paper, which help in extending the SP framework to SISO OTFS systems,  are summarized below.
{\begin{itemize}
\item We propose SP-based  OTFS framework which, as shown in Fig.~\ref{fig:Frame}(b), superimposes pilots on to data symbols. We consider a  frame which, unlike the frame  of  \cite{DBLP:journals/tvt/RavitejaPH19}  in Fig.~\ref{fig:Frame}(a), does not i) insert zeros between data and pilot symbols; and ii) require dedicated delay-Doppler slot to transmit pilots.  The proposed framework, unlike \cite{DBLP:conf/globecom/RamachandranC18}, does not require a dedicated pilot frame to estimate channel.{The proposed framework, will thus have significantly higher SE than \cite{DBLP:journals/tvt/RavitejaPH19, DBLP:conf/globecom/RamachandranC18}, with a minor BER degradation, a fact we will numerically validate later.} 

\item  We propose two  SP-aided channel estimation and data detection designs --  SP-non-iterative (SP-NI) and SP-iterative (SP-I). The SP-NI design exploits OTFS channel sparsity by performing minimum mean square error (MMSE) channel estimation in the delay-Doppler domain. It then detects data using  computationally-efficient message passing algorithm \cite{raviteja2018interference}, which again exploits the delay-Doppler domain channel sparsity.  Its BER and SE, however, due to the mutual interference between data and pilots, degrades  at high signal to noise ratio (SNR) values. The SP-I design mitigates this interference  by iterating between channel estimation and data detection in the delay-Doppler domain, and has better BER and SE. 

\item We also derive a lower bound on the  signal-to-interference-plus-noise-ratio (SINR) of the  two proposed designs. This lower bound is then maximized by deriving a closed-form expression to optimally allocate transmit power between data and pilot symbols.  {We show that the optimal power minimizes the BER and  maximizes the SE of both the designs.}

\item {We numerically i) validate the tightness of the theoretical MSE and the optimal power allocation; and  ii) investigate the effect of power distribution between data and pilot symbols on the MSE, BER, and SE of the proposed designs.  
We also numerically show that the proposed SP-I design, to achieve a SE of $3$ bps/Hz, require $\approx 15$ dB and $\approx 5$ dB lower SNR than the designs in  and \cite{DBLP:conf/globecom/RamachandranC18} and \cite{DBLP:journals/tvt/RavitejaPH19}, respectively.} 
\end{itemize}}


\textbf{Notations:} Lower and upper case bold face letters $\mathbf{a}$ and $\mathbf{A}$ denote vectors and matrices, respectively. The superscript $(\cdot)^H$ denotes Hermitian
transposition operator and the operation $\mathbf{A}\otimes \mathbf{B}$ denotes  Kronecker product of the matrices $\mathbf{A}$ and~$\mathbf{B}$. The operator $\text{vec}(\mathbf{A})$ vectorises the matrix $\mathbf{A}$, and $\mathbb{E}\{\cdot\}$ represents the exception operation. The operation $\mathbf{A}\left(p,q\right)$ extracts the $(p,q)$th element of the matrix $\mathbf{A}$. The notation $\text{diag}\{a_0,a_1,\ldots,a_{N-1}\}$ denotes an $N\times N$ diagonal matrix. The notations $\mathbf{I}_{N}$ and  $\mathbf{0}_{M\times N}$ represents $N\times N$ identity and zero matrices, respectively. The operation $\mbox{Tr}(\mathbf{A})$ computes the trace of the matrix $\mathbf{A}$ and $ \left[.\right]_{M} $ denotes modulo $ M $ operation. The operators  $\mbox{Var}[x]$ and $\mathbb{E}_{\mathbf{a}|\mathbf{b}}$ define the variance and conditional expectation operations, respectively.

%
%

%
%

  \vspace{-0.4cm}
\section{OTFS system model with superimposed pilot}
\label{system}
\subsection{Transmit vector generation}
We consider, as shown in  Fig.~\ref{fig:BD}, a {single-antenna} SP-OTFS system which transmits a data symbol $x_{d}\left[l,k \right] $  at the $ l $th delay and the $ k $th Doppler location, with $ l=0,1, \cdots, M-1$ and  $k=0,1, \cdots N-1$. Here $ M$ and $N$ denote the number of {delay and Doppler bins}, respectively. 
			\vspace{-0.5cm}
\begin{figure}[htbp]
	\centering
	{\includegraphics[scale=0.65]{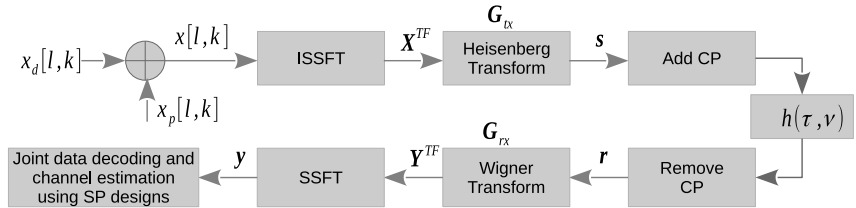}}
	\caption{\small OTFS system model with superimposed pilot (SP) sequence.}
	\label{fig:BD}
			\vspace{-0.7cm}
\end{figure}
The data symbol $x_{d}\left[l,k \right] $   is superimposed on to  the pilot symbol $x_{p}\left[l,k \right] $ in the delay-Doppler domain by arithmetical addition as follows
 \begin{equation}
 \label{eq:trsymbol}
 x\left[l,k \right]=x_{d}\left[l,k\right]+x_{p}\left[l,k\right].
 \end{equation}
 Equation \eqref{eq:trsymbol} is next re-expressed in matrix form as 
\begin{equation} \label{eq:trsymbol_mtx}
\mathbf{X}=\mathbf{X}_{d}+\mathbf{X}_{p},
\end{equation} 
by arranging  $x_{d}\left[l,k\right]$, $x_{p}\left[l,k\right]$, and $x\left[l,k\right]$ in  matrix form as $ \mathbf{X}_{d}\in \mathbb{C}^{M\times N} $, $ \mathbf{X}_{p}\in \mathbb{C}^{M\times N} $ and $ \mathbf{X}\in \mathbb{C}^{M \times N} $, respectively. 
The zero-mean independent and identically distributed (i.i.d.) elements of the matrices $\mathbf{X}_{d}$ and $\mathbf{X}_{p}$ are assumed to have   $ \mathbb{E}\big\{\left|x_{d}\left[l,k \right] \right|^{2} \big\} = \sigma_{d}^{2}$ and $ \mathbb{E}\big\{\left|x_{p}\left[l,k \right] \right|^{2} \big\} = \sigma_{p}^{2}$, respectively. We also impose the following constraint that $ \mathbb{E}\big\{\left|x\left[l,k \right] \right|^{2} \big\}=\sigma_{d}^{2} + \sigma_{p}^{2}=1 $. 

The OTFS scheme, as shown in Fig.~\ref{fig:BD}, first maps the $MN$ symbols in the delay-Doppler domain to the time-frequency domain by using ISFFT.
Let $\mathbf{X}^{\text{TF}}\in \mathbb{C}^{M\times N}$ be the superimposed symbol matrix in the time-frequency domain whose $(m,n)$th element is given as
\begin{equation}
\label{eqn:nm}
x^{\text{TF}}\left[m,n \right]=\dfrac{1}{\sqrt{NM}}\sum_{l=0}^{M-1}\sum_{k=0}^{N-1}x\left[l,k\right] e^{j2\pi\Big(\dfrac{ml}{M}-\dfrac{nk}{N} \Big)},
\end{equation}
  where $ m=0,1, \ldots, M-1$ and  $n=0,1, \ldots, N-1$. The time-frequency frame has a duration $NT$, and  bandwidth $M\Delta f$. Here $T$ and $\Delta f$, with $ T\Delta f=1 $ \cite{raviteja2018interference}, are the sampling intervals along the time and frequency axis, respectively.  The time-frequency symbol matrix $ \mathbf{X}^{\text{TF}}$, using \eqref{eq:trsymbol_mtx} and \eqref{eqn:nm}, can be represented as the function of the delay-Doppler matrix $\mathbf{X}$ as $ \mathbf{X}^{\text{TF}}=\mathbf{F}_{M}\mathbf{X}\mathbf{F}_{N}^{H} $  \cite{raviteja2018practical}.
Here $ \mathbf{F}_{M}\in \mathbb{C}^{M\times M} $ and $ \mathbf{F}_{N}\in \mathbb{C}^{N \times N} $ are the normalized discrete Fourier transform (DFT) matrices with $\mathbf{F}_{M}\left(p,q\right)= \big(1/\sqrt{M} \big)\exp \left(-j2\pi pq/M \right)$ and $\mathbf{F}_{N}\left(p,q\right)= \big(1/\sqrt{N} \big)\exp \left(-j2\pi pq/N \right)$.  The time-frequency domain samples $x^{\text{TF}}[m,n]$ are pulse-shaped using a transmit pulse $g_{\text{tx}}(t)$ to generate a continuous-time signal $s(t)$ by using the Heisenberg transform \cite{raviteja2018interference}. The  signal $s(t)$, sampled at a rate $f_s=M\Delta f=M/T$, can be expressed in the matrix form  as \cite{raviteja2018practical}
\begin{eqnarray}\label{eq:S_Mtx}
\mathbf{S}=\mathbf{G}_{\text{tx}}\mathbf{F}_{M}^{H}\mathbf{X}^{\text{TF}} 
= \mathbf{G}_{\text{tx}}\mathbf{X}\mathbf{F}_{N}^{H}.
\end{eqnarray}
The matrix $\mathbf{S}\in\mathbb{C}^{M\times N}$ consists of $MN$ samples of the signal $s(t)$, and the diagonal matrix $ \mathbf{G}_{\text{tx}} \in \mathbb{C}^{M\times M} $ is obtained by sampling the transmit pulse $ g_{\text{tx}}\left(t\right)$ at the time instants $\frac{mT}{M}$ with $m=0,1,\ldots,M-1$. The time-domain transmit vector  $\mathbf{s}$, shown  in Fig.~\ref{fig:BD}, is derived as $ \mathbf{s}=\mbox{vec}\left(\mathbf{S}\right) \in \mathbb{C}^{MN \times 1} $. By substituting $\mathbf{X}$ from \eqref{eq:trsymbol_mtx}, the vector $ \mathbf{s} $ can be expressed using the identity $\mbox{vec}(\mathbf{A}\mathbf{B}\mathbf{C})=\big(\mathbf{C}^T\otimes \mathbf{A}\big)\mbox{vec}(\mathbf{B})$ \cite{petersen2012matrix} {and the property $\mathbf{F}_N^T=\mathbf{F}_N$ of the DFT matrix $\mathbf{F}_N$} as 
\begin{eqnarray}
\label{eqn:ss}
\mathbf{s}= \left(\mathbf{F}_{N}^{H} \otimes \mathbf{G}_{\text{tx}} \right)\mathbf{x}.
\end{eqnarray} 
Here $ \mathbf{x}=\mbox{vec}\left(\mathbf{X}\right) = \mathbf{x}_{p}+\mathbf{x}_{d} $ with $\mathbf{x}_{d}\in\mathbb{C}^{NM\times 1}$ and $\mathbf{x}_{p}\in\mathbb{C}^{NM\times 1}$ being the data and symbol vectors in the delay-Doppler domain, respectively.
To mitigate inter-frame interference in the time domain, as shown in Fig.~\ref{fig:BD}, a cyclic prefix (CP) of length of $ l_{max } $ samples is appended to the transmit signal vector $ \mathbf{s} $, where $ l_{max} $ is the tap corresponding to the maximum delay $ \tau_{max} $. 
  \vspace{-0.6cm}
\subsection{Channel impulse response}
{A  delay-Doppler domain channel coefficient is characterized by a cluster, which ideally consists of infinite many reflectors. We assume that there are $Q$ such clusters in the channel, where the $i$th cluster has a dominant component $h_i$ with associated delay $\tau_i$ and Doppler shift $\nu_i$ \cite{DBLP:journals/twc/SurabhiAC19}.} The impulse response of the wireless channel $h\left(\tau , \nu \right)$ in the delay-Doppler domain is \cite{raviteja2018interference}
\begin{equation}
h\left(\tau , \nu \right)= \sum_{i=1}^{Q} h_{i} \delta \left(\tau - \tau_{i}\right) \delta\left(\nu - \nu_{i}\right).
\end{equation}
Here $ \tau \in \left[0, \tau_{max} \right] $ and $ \nu \in \left[-\nu_{max}, \nu_{max} \right] $ are the delay and Doppler shifts respectively, with $\tau_{max}$ and $ \nu_{max} $ being the maximum delay and the maximum Doppler shift among all channel paths. The quantity $Q$ is the number of propagation paths, $ h_{i} $ is the complex path gain for the $i$th path, which is distributed as $\mathcal{CN}(0,\sigma_{h_{i}}^{2})$. The delay $ \tau_{i} $ and the Doppler $ \nu_{i} $, associated with the $ i $th path, are expressed as $\tau_{i}=\frac{l_{i}}{M\Delta f}$ and $\nu_{i}=\frac{k_{i}}{NT}$, respectively. {Here the integers $l_{i}$ and $k_{i}$ respectively denote the delay and Doppler taps for the $i$th path.} We, similar to \cite{raviteja2018interference}, assume $l_i$ to be an integer. This is because in a typical wide-band system, the  sampling time resolution $\frac{1}{M\Delta f}$ is sufficient to approximate the path delays to the nearest sampling point \cite{raviteja2018interference}. For the sampling rate $f_{s}=M/T $, the maximum channel delay is assumed to be $ \tau_{max}=\left(l_{max }\right)T/M $, which implies that $ l_{i}\in \left[0,l_{max }\right] $. We also, similar to \cite{DBLP:journals/twc/SurabhiAC19}, do not consider the effect of fractional Doppler. {This is because  the Doppler resolution $ 1/NT $  progressively reduces with increasing number of Doppler bins $ N $ \cite{DBLP:journals/twc/SurabhiAC19}}. The proposed designs, however, can be easily extended to fractional Doppler case.     

\subsection{Receive processing}
The discrete baseband received signal vector $ \mathbf{r}\in \mathbb{C}^{MN \times 1} $,  after removing the CP,   is  \cite{raviteja2018practical} 
$\mathbf{r}=\mathbf{H}\mathbf{s} + \mathbf{w}$. The noise $ \mathbf{w}\in \mathbb{C}^{MN \times 1} $ has  i.i.d. complex Gaussian entries with zero mean  and variance $ \sigma_{w}^{2} $. The channel $ \mathbf{H}\in \mathbb{C}^{MN \times MN} $ is \emph{sparse}, and is  $\mathbf{H}=\sum_{i=1}^{Q} h_{i} \boldsymbol{\Pi}^{l_{i}}\boldsymbol{\Delta}^{k_{i}}$\cite{raviteja2018practical}. Here
\begin{footnotesize}
\begin{equation}
\boldsymbol{\Pi} \small\overset{\Delta}{=} \begin{bmatrix}
0 & \cdots & 0 & 1 \\ 
1 & \cdots & 0 & 0 \\ 
\vdots & \ddots & \vdots & \vdots \\ 
0 & \cdots & 1 & 0 
\end{bmatrix}.
\end{equation}
\end{footnotesize}
is the $ MN \times MN $ forward cyclic-shift (permutation) matrix, and  
$\boldsymbol{\Delta}=\mbox{diag}\left\{z^{0}, z^{1}, \ldots, z^{MN-1} \right\} 
$, is a diagonal matrix. 
Also,  $ \small z=\mbox{exp}\left(2\pi j/MN \right) $ with $j=0,1,\ldots,MN-1$. {Note that the matrices $ \boldsymbol{\Pi} $ and $ \boldsymbol{\Delta} $ model the delays and Doppler shifts, respectively.}       
The time-frequency domain receive signal matrix $ \mathbf{Y}^{\text{TF}} \in \mathbb{C}^{M\times N} $,  as shown in Fig.~\ref{fig:BD}, is derived from the received signal $\mathbf{r}$ using the Wigner transformation (inverse of Heisenberg transform) as $ \mathbf{Y}^{\text{TF}}=\mathbf{F}_{M}\mathbf{G}_{\text{rx}}\mathbf{R} $ \cite{raviteja2018practical}. 
Here $\mathbf{R}=\mbox{vec}^{-1}(\mathbf{r})\in \mathbb{C}^{M\times N}$, and the diagonal matrix $ \mathbf{G}_{\text{rx}}\in\mathbb{C}^{M\times M} $ is obtained by sampling the receive pulse $ g_{\text{rx}}\left(t\right) $  at the time instants $\frac{mT}{M}$ with $m=0,1,\ldots,M-1$. The delay-Doppler receive signal matrix $ \mathbf{Y}\in \mathbb{C}^{M\times N} $ is obtained from $ \mathbf{Y}^{\text{TF}} $ using the SFFT operation as \cite{raviteja2018practical}
\begin{eqnarray}\label{eq:otfs_rec1}
\mathbf{Y}=\mathbf{F}_{M}^{H}\mathbf{Y}^{\text{TF}}\mathbf{F}_{N} 
= \mathbf{G}_{\text{rx}}\mathbf{R}\mathbf{F}_{N}.
\end{eqnarray}  
The receive vector $\mathbf{y}=\mbox{vec}\left(\mathbf{Y}\right)$, using the identity $\mbox{vec}(\mathbf{A}\mathbf{B}\mathbf{C})=\big(\mathbf{C}^T\otimes \mathbf{A}\big)\mbox{vec}(\mathbf{B})$ \cite{petersen2012matrix}, is 
\begin{eqnarray}
\label{eqn:rxvectordd}
\mathbf{y}=\left(\mathbf{F}_{N} \otimes \mathbf{G}_{\text{rx}} \right) \mathbf{r}  = \left(\mathbf{F}_{N} \otimes \mathbf{G}_{\text{rx}} \right) \left(\mathbf{H}\mathbf{s}+ \mathbf{w} \right).
\end{eqnarray}
By substituting (\ref{eqn:ss}) in (\ref{eqn:rxvectordd}), we get 
\begin{eqnarray}
\label{eqn:rxvectr12}
\mathbf{y}=\left( \mathbf{F}_{N} \otimes \mathbf{G}_{\text{rx}} \right)\mathbf{H} \left( \mathbf{F}_{N}^{H} \otimes \mathbf{G}_{\text{tx}} \right)\mathbf{x} + \tilde{\mathbf{w}}  = \mathbf{H}_{\text{eff}}\left(\mathbf{x}_{p}+\mathbf{x}_{d}\right)+ \tilde{\mathbf{w}},
\end{eqnarray}
where the vector $ \tilde{\mathbf{w}}= \left(\mathbf{F}_{N} \otimes \mathbf{G}_{\text{tx}} \right)\mathbf{w} $, and the effective channel matrix $ \mathbf{H}_{\text{eff}} \in \mathbb{C}^{MN \times MN} $ is 
\begin{equation}\label{eq:eff_chan}
\mathbf{H}_{\text{eff}}=\left( \mathbf{F}_{N} \otimes \mathbf{G}_{\text{rx}} \right)\mathbf{H} \left( \mathbf{F}_{N}^{H} \otimes \mathbf{G}_{\text{tx}} \right).
\end{equation}
If both transmitter and receiver use  rectangular pulse shapes of duration $T$, we have $ \mathbf{G}_{\text{rx}}=\mathbf{G}_{\text{tx}}=\mathbf{I}_{M} $. {The resulting inter-symbol-interference (ISI) and inter-carrier-interference (ICI) in time-frequency domain can be included in the effective channel matrix $\mathbf{H}_{\text{eff}}$ as  \cite{raviteja2018interference, raviteja2018practical}}:
\begin{align}
\label{eqn:heff}
\mathbf{H}_{\text{eff}}
= \left(\mathbf{F}_{N} \otimes \mathbf{I}_{M} \right) \left(\sum_{i=1}^{Q} h_{i} \boldsymbol{\Pi}^{l_{i}}\boldsymbol{\Delta}^{k_{i}} \right) \left(\mathbf{F}_{N}^{H} \otimes \mathbf{I}_{M} \right) = \mathbf{B}_{\text{rx}}\left(\sum_{i=1}^{Q} h_{i}\boldsymbol{\Theta}_{i} \right)\mathbf{B}_{\text{tx}}.
\end{align} 
Here the matrices $\mathbf{B}_{\text{tx}}=\left(\mathbf{F}_{N}^{H} \otimes \mathbf{I}_{M} \right)$, $\mathbf{B}_{\text{rx}}=\left(\mathbf{F}_{N} \otimes \mathbf{I}_{M} \right)$ and $\boldsymbol{\Theta}_{i}=\left(\boldsymbol{\Pi}^{l_{i}}\boldsymbol{\Delta}^{k_{i}}\right)$ are of size $MN\times MN$. {We see that the OTFS channel model is radically different from its OFDM counterpart, which significantly changes the signal processing for SP-aided transmission in OTFS systems.} The received signal vector $\mathbf{y}$, by substituting $ \mathbf{H}_{\text{eff}} $ from (\ref{eqn:heff}) in (\ref{eqn:rxvectr12}),  can be recast as follows
\begin{eqnarray}
\label{eqn:inopy1}
\mathbf{y}=\left(\sum_{i=1}^{Q} h_{i}\boldsymbol{\Gamma}_{i} \right)\left(\mathbf{x}_{p}+\mathbf{x}_{d} \right) + \tilde{\mathbf{w}}.
\end{eqnarray}
The matrix $ \boldsymbol{\Gamma}_{i} \in \mathbb{C}^{MN \times MN}$ is given as $\boldsymbol{\Gamma}_{i}=\mathbf{B}_{\text{rx}}\boldsymbol{\Theta}_{i}\mathbf{B}_{\text{tx}}$. The received vector in \eqref{eqn:inopy1} for the SP-based OTFS system can be succinctly rewritten as
\begin{eqnarray}
\label{eqn:inopy}
\mathbf{y}=\boldsymbol{\Omega}_{p}\mathbf{h} + \boldsymbol{\Omega}_{d}\mathbf{h}+\tilde{\mathbf{w}}.
\end{eqnarray}
The concatenated matrices $\boldsymbol{\Omega}_{p}\in\mathbb{C}^{MN \times  Q}$ and $\boldsymbol{\Omega}_{d}\in\mathbb{C}^{MN \times  Q}$ corresponding to the pilot vector $\mathbf{x}_p$ and data vector $\mathbf{x}_d$ are obtained as follows
\begin{eqnarray}
\label{eqn:pilot}
\boldsymbol{\Omega}_{p}&=& \left[\boldsymbol{\Gamma}_{1}\mathbf{x}_{p} \; \; \boldsymbol{\Gamma}_{2}\mathbf{x}_{p} \;  \; \cdots \;  \; \boldsymbol{\Gamma}_{Q}\mathbf{x}_{p} \right] \\
\label{eqn:data}
\boldsymbol{\Omega}_{d}&=& \left[\boldsymbol{\Gamma}_{1}\mathbf{x}_{d} \; \; \boldsymbol{\Gamma}_{2}\mathbf{x}_{d} \;  \; \cdots \; \; \boldsymbol{\Gamma}_{Q}\mathbf{x}_{d} \right].
\end{eqnarray}
{The matrix $\boldsymbol{\Gamma}_{i}$ in the $\mathbf{\Omega}_d$ and $\mathbf{\Omega}_p$, as observed from Appendix~\ref{Lamma_proof} and Appendix~\ref{Appendix_D}, significantly complicates their mathematical characterization.} The delay-Doppler domain channel vector $ \mathbf{h}=\left[h_{1}, h_{2}, \cdots, h_{Q}\right]^T\in \mathbb{C}^{Q\times 1} $ has  zero mean and covariance matrix
\begin{eqnarray}
\label{Covh1}
\mathbf{C}_{h}=\mathbb{E}\left\{\mathbf{h}\mathbf{h}^{H} \right\}=\mbox{diag}\left\{\sigma^{2}_{h_{1}}, \cdots, \sigma^{2}_{h_{Q}}  \right\}.
\end{eqnarray}
We next state the following result related to random matrix $ \boldsymbol{\Omega}_{d}$, which is proved in Appendix-\ref{Lamma_proof}.
 \vspace{-0.7cm}
{\begin{lemma}\label{Lemma1}
The random matrix $ \boldsymbol{\Omega}_{d}$ has the following statistical characteristics
		\begin{eqnarray}
		\label{eqn:Omegaexp1}
		\mathbb{E}\left\{\boldsymbol{\Omega}_{d}\right\}=\boldsymbol{0}_{MN\times Q}\ \ \text{and}\ \ \
		\mathbb{E}\left\{\boldsymbol{\Omega}_{d}\boldsymbol{\Omega}_{d}^{H} \right\}=\sigma_{d}^{2}Q\mathbf{I}_{MN}.
		\end{eqnarray}
	\end{lemma}
The noise vector $ \tilde{\mathbf{w}}=\left(\mathbf{F}_{N} \otimes \mathbf{I}_{M} \right)\mathbf{w}$ of size ${MN \times 1}$ has
\begin{eqnarray}
\label{eqn:Covnoise}
\mathbb{E}\left\{\tilde{\mathbf{w}}\right\}=\boldsymbol{0}_{MN \times 1}\ \ \text{and}\ \ \  
\mathbf{C}_{\tilde{\mathbf{w}}}=\mathbb{E}\left\{\tilde{\mathbf{w}}\tilde{\mathbf{w}}^{H} \right\}= \sigma_{w}^{2}\mathbf{I}_{MN}.
\end{eqnarray}
 \vspace{-0.8cm}
\begin{remark}
The input-output relations in \eqref{eqn:rxvectr12} and \eqref{eqn:inopy} are same, with different mathematical formulations. They will, therefore, be used interchangeably  for developing the proposed designs.
\end{remark}

 \vspace{-0.6cm}
\section{Superimposed pilot -based OTFS channel estimator and data detector}\label{IDT_and_CE}
This section proposes the SP-NI and SP-I designs for estimating channel and detecting data in OTFS systems. We begin with the SP-NI design which calculates the MMSE channel estimate in the delay-Doppler domain, by treating data as interference. {It later detects data using the reduced-complexity message passing algorithm \cite{raviteja2018interference}, which exploits the delay-Doppler domain channel sparsity. This algorithm, with a significantly lower computational complexity, yields a BER which is very close to the  maximum-a-posterior probability (MAP) detector \cite{raviteja2018interference}.
  \vspace{-0.6cm}
\subsection{Proposed superimposed pilot non-iterative (SP-NI) design} \label{Pilot_aided_CE}
 \vspace{-0.2cm}
The proposed SP-NI design, to exploit the fact that very few parameters are needed to model the delay-Doppler channel, estimates it in the same domain by considering data as an interference. Furthermore, to incorporate message passing algorithm, which  exploits the OTFS channel sparsity  for data detection, the proposed design evaluates both first- and second-order statistical characteristics of the SP-aided channel estimation error. The statistics  are then used to calculate the probability mass functions (pmfs) for the message passing algorithm.
 \vspace{-0.2cm}
\subsubsection{Channel estimation in SP-NI design}
The received signal vector in (\ref{eqn:inopy}) is rewritten as 
\begin{equation}
\label{eqn:receiver_obs1}
\mathbf{y}=\boldsymbol{\Omega}_{p}\mathbf{h}+\underbrace{\boldsymbol{\Omega}_{d}\mathbf{h}+\tilde{\mathbf{w}}}_{\tilde{\mathbf{w}}_{d}}=\boldsymbol{\Omega}_{p}\mathbf{h}+\tilde{\mathbf{w}}_{d}.
\end{equation}
The mean $ \boldsymbol{\mu}_{\tilde{\mathbf{w}}_{d}} $ and covariance matrix $ \mathbf{C}_{\tilde{\mathbf{w}}_{d}} $ of the noise-plus-interference vector $\tilde{\mathbf{w}}_{d}$ are derived in the following lemma, whose proof is relegated to Appendix-\ref{Lamma_proof_2}.  
\begin{lemma}\label{Lemma2}
Since $ \mathbb{E}\left\{\boldsymbol{\Omega}_{d} \right\}=\boldsymbol{0}_{MN\times Q} $ and $\mathbb{E}\left\{\mathbf{h} \right\}=\boldsymbol{0}_{Q\times 1}$, the mean of the vector $ \tilde{\mathbf{w}}_d $ is
\begin{equation}
\boldsymbol{\mu}_{\tilde{\mathbf{w}}_{d}}=\mathbb{E}\left\{\tilde{\mathbf{w}}_{d}\right\}=\boldsymbol{0}_{MN\times 1}
\end{equation}
and its covariance matrix using \emph{Lemma}~\ref{Lemma1} is 
\begin{eqnarray}
\label{eqn:covariance1}
\mathbf{C}_{\tilde{\mathbf{w}}_{d}}=\mathbb{E}\left\{\tilde{\mathbf{w}}_{d}\tilde{\mathbf{w}}_{d}^{H} \right\} = \left( \left(\sum_{i=1}^{Q}\sigma^{2}_{h_{i}} \right)\sigma_{d}^{2} + \sigma_{w}^{2}  \right)\mathbf{I}_{MN}.
\end{eqnarray}
\end{lemma}
{The proof also uses the following lemma  from \cite{SinghMJV19,mishra2017affine}}.
\begin{lemma}\label{Lemma21}
	If  $ \mathbf{A}\in \mathbb{C}^{m \times n} $ is a random matrix such that  $\mathbb{E}\left\{\mathbf{A}\mathbf{A}^{H}\right\}=\sigma^{2}\mathbf{I}_{m}  $, then for any hermitian matrix $ \mathbf{B} \in \mathbb{C}^{n \times n} $, $ \mathbb{E}\left\{\mathbf{A}\mathbf{B}\mathbf{A}^{H}\right\}=\dfrac{\text{Tr}\left(\mathbf{B}\right)}{n}\mathbb{E}\left\{\mathbf{A}\mathbf{A}^{H}\right\} $. 
\end{lemma}
The MMSE estimate  $ \hat{\mathbf{h}}_{\text{NI }}$ of the channel vector $ \mathbf{h} $ using the SP-NI design and the  linear observation model in (\ref{eqn:receiver_obs1}), is given as \cite{SinghBV19} 
 \begin{equation}
 \label{eqn:hathinitial33}
 \hat{\mathbf{h}}_{\text{NI }}= \left(\boldsymbol{\Omega}_{p}^{H}\mathbf{C}_{\tilde{\mathbf{w}}_{d}}^{-1}\boldsymbol{\Omega}_{p}+ \mathbf{C}_{\mathbf{h}}^{-1}\right)^{-1}\boldsymbol{\Omega}_{p}^{H}\mathbf{C}_{\tilde{\mathbf{w}}_{d}}^{-1}\mathbf{y}.
 \end{equation}
The covariance matrices $ \mathbf{C}_{\mathbf{h}}$ and $\mathbf{C}_{\tilde{\mathbf{w}}_{d}} $ are given in   (\ref{Covh1}) and (\ref{eqn:covariance1}), respectively. {\textit{We observe that the channel estimator in \eqref{eqn:hathinitial33} exploits OTFS channel  sparsity by computing the  inverse of a $Q\times Q$ matrix, where $Q\ll MN$ denotes the number delay-Doppler taps.}} The error covariance matrix of $\hat{\mathbf{h}}_{\text{NI }}$ can next be calculated as \cite{kay1993fundamentals} 
\begin{eqnarray}
\label{eqn:incov}
\boldsymbol{\Sigma}_{\text{NI }}
= \left(\boldsymbol{\Omega}_{p}^{H}\mathbf{C}^{-1}_{\tilde{\mathbf{w}}_d} \boldsymbol{\Omega}_{p}+ \mathbf{C}^{-1}_{\mathbf{h}}\right)^{-1}.
\end{eqnarray} 
{We see that, unlike pilot-aided channel estimation methods in \cite{DBLP:journals/tvt/RavitejaPH19,DBLP:conf/globecom/RamachandranC18}, the covariance matrix $\mathbf{C}_{\tilde{\mathbf{w}}_d}$ in the proposed design, due to superimposed pilots, includes the effect of mutual interference between data and pilot symbols.} 
The mean squared error (MSE) of channel estimate is 
\begin{eqnarray}
\label{eqn:MSEin}
B_{h, \text{NI}}=\mathbb{E}\big\{\|\mathbf{h}-\hat{\mathbf{h}}_{\text{NI }} \|^{2} \big\}= \mbox{Tr}\left[ \boldsymbol{\Sigma}_{\text{NI }}\right].
\end{eqnarray}
After estimating channel using superimposed pilot symbols, we next employ message passing algorithm  to detect the data vector $\mathbf{x}_d$ \cite{raviteja2018interference}.
\subsubsection{Message passing-aided data detection in the SP-NI design}
 The low-complexity message passing algorithm  exploits sparsity of the delay-Doppler OTFS channel while detecting data symbols \cite{DBLP:conf/wcnc/RavitejaPJHV18, raviteja2018interference}. The signal used for detecting data is obtained by subtracting the pilot signal from the received signal  in (\ref{eqn:rxvectr12}) as follows
\begin{eqnarray} 
\label{eqn:yd1}
\mathbf{y}_{d}=\mathbf{y}-\widehat{\mathbf{H}}_{\text{eff-NI}}\mathbf{x}_{p} 
= \mathbf{H}_{\text{eff}}\mathbf{x}_{d}+\tilde{\mathbf{w}}_e.
\end{eqnarray}
{The vector $\tilde{\mathbf{w}}_e=\mathbf{e}_{p}+\tilde{\mathbf{w}}$ consists of channel estimation error and noise, with the error vector  
\begin{equation}
\mathbf{e}_{p}=\left(\mathbf{H}_{\text{eff}}-\widehat{\mathbf{H}}_{\text{eff-NI}} \right)\mathbf{x}_{p}.
\end{equation}
The estimate of  effective channel matrix ${\mathbf{H}}_{\text{eff}} $, defined in \eqref{eqn:heff}, is  obtained  by substituting the expression of channel estimate $ \hat{\mathbf{h}}_{\text{NI }} $  in \eqref{eqn:heff} as 
 \begin{equation}\label{eq:Heff_impract}
\widehat{\mathbf{H}}_{\text{eff-NI}}=\mathbf{B}_{\text{rx}}\left(\sum_{i=1}^{Q}\hat{h}_{\text{NI},i}\boldsymbol{\Theta}_{i}\right)\mathbf{B}_{\text{tx}}.
\end{equation}}
\begin{figure}
	\centering
	{\includegraphics[scale=0.7]{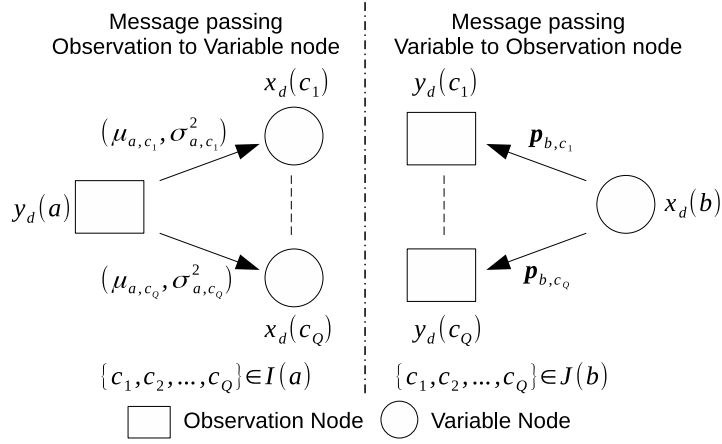}}
		 \vspace{-0.3cm}
	\caption{\small Graphical representation of the message passing algorithm for data detection in the SP-NI design.}
	\label{fig:MPalg}
	 \vspace{-1cm}
\end{figure}
{We now use the system model in \eqref{eqn:yd1} for developing the message-passing  receiver. To begin with, let $\mathcal{I}(a)$ and $\mathcal{J}(b)$ be the sets containing indices of non-zero elements in the $a$th row and the $b$th column of the matrix $\widehat{\mathbf{H}}_{\text{eff-NI}}$, respectively.  This  implies that $|\mathcal{I}(a)|=|\mathcal{J}(b)|=Q$. Figure~\ref{fig:MPalg} shows the factor-graph for the message-passing detection algorithm which consists of observation and variable nodes \cite{SomDSCR11}, which are defined next. Before doing that, it is worth noting that due to the OTFS channel sparsity, an observation node $y_d(a)$, for $1\leq a\leq MN$, is connected to only $Q$ number of variable nodes $x_d(c_1),\ldots,x_d(c_Q)$, where $Q\ll MN$ and $\{c_1,\ldots,c_Q\}\in \mathcal{I}(a)$. Similarly, a variable $x_d(b)$, for $1\leq b\leq MN$, is connected to $Q$ number of observation nodes $y_d(c_1),\ldots,y_d(c_Q)$, where $\{c_1,\ldots,c_Q\}\in \mathcal{J}(b)$.} 

We now use \eqref{eqn:yd1} to write the input-output relation for the $a$th observation node $\mathbf{y}_d (a)$ with the $b$th variable nodes $\mathbf{x}_d(b)$, for $b\in\mathcal{I}(a)$  as
\begin{align}
\mathbf{y}_{d}\left(a\right)=\mathbf{x}_{d}\left(b\right)\mathbf{H}_{\text{eff}}\left(a,b\right)+ \gamma_{a,b},
\end{align}
where $1\leq a\leq MN$ and the interference-plus-noise term 
\begin{equation}\label{eq:Gamma_intr}
\gamma_{a,b}= \sum_{c \in \mathcal{I}\left(a\right), c\neq b} \mathbf{x}_{d}\left(c\right)\mathbf{H}_{\text{eff}}\left(a,c\right)+\tilde{\mathbf{w}}_{e}\left(a\right).
\end{equation}
{The first part of the expression, by using central-limit theorem \cite{raviteja2018interference}, can be approximated as a Gaussian random variable. Since $\tilde{\mathbf{w}}_e$, defined below \eqref{eqn:yd1}, is also Gaussian distributed, the quantity $\gamma_{a,b}$ can also be approximated as a Gaussian random variable. The message passing algorithm,  in the $i$th iteration,  passes the messages in terms of mean $\mu^{(i)}_{a,b}$ and variance $\big(\sigma^{(i)}_{a,b}\big)^2$ of the random parameter $\gamma_{a,b}$, from the observation node $y_d(a)$ to the variable node $x_d(b)$, where $b\in \mathcal{I}(a)$. 
	\begin{remark}
	We crucially note that, in contrast to \cite{DBLP:journals/tvt/RavitejaPH19},  the interference-plus-noise term $\gamma_{a,b}$ is a  function of $\tilde{\mathbf{w}}_e$ which, as shown in \eqref{eqn:yd1}, is a function of the noise  $\tilde{\mathbf{w}}$ and the interference  $\mathbf{e}_p$, which is due to imperfect channel estimate. The superimposed pilots are only partially canceled  while detecting data. {This is unlike the  message passing i) algorithm in \cite{raviteja2018interference}, which assumes perfect channel; ii) receiver in \cite{DBLP:journals/tvt/RavitejaPH19} which, due to insertion of zeros, need not consider the mutual interference between data and pilot symbols. {The mutual interference in the proposed SP-aided designs significantly changes the message calculations.} We have to, in contrast,  derive the first- and second-order statistical characteristics of the vector $\mathbf{e}_p$ for evaluating the mean $\mu^{(i)}_{a,b}$ and the variance $\big(\sigma^{(i)}_{a,b}\big)^2$ of $\gamma_{a,b}$.} {We see from \eqref{eq:Heff_impract} that the estimated channel matrix $\widehat{\mathbf{H}}_{\text{eff-NI}}$ in the $\mathbf{e}_p$ expression is a function of the estimate $\hat{h}_{\text{NI},i}$ and matrices $\mathbf{B}_{\text{tx}}$, $\mathbf{B}_{\text{rx}}$, $\mathbf{\Theta}_{i}$. The estimate $\hat{h}_{\text{NI},i}$, as shown in  \eqref{eqn:hathinitial33}, is a function of pilot matrix $\boldsymbol{\Omega}_p$, covariance matrix $\mathbf{C}_{\tilde{\mathbf{w}}_d}$ of  data interference  and the observation vector $\mathbf{y}$. The statistical characterization of $\mathbf{e}_p$, as shown next, is thus non-trivial.} 
\end{remark}

Since $ \mathbb{E}\left\{ h_{i}\right\}=0 $, it follows from \eqref{eqn:heff} that $\mathbb{E}\left\{\mathbf{H}_{\text{eff}} \right\}=\boldsymbol{0}_{MN\times MN}$. We, consequently, have  from (\ref{eqn:rxvectr12}), $ \mathbb{E}\left\{\mathbf{y} \right\}=\boldsymbol{0}_{MN\times 1} $, which also implies $ \mathbb{E}\big\{\hat{\mathbf{h}}_{\text{NI }}\big\}=\boldsymbol{0}_{Q\times 1}$. We finally have, by using the property $ \mathbb{E}\big\{\hat{\mathbf{h}}_{\text{NI }}\big\}=\boldsymbol{0}_{Q\times 1}$, that  $\mathbb{E}\big\{\widehat{\mathbf{H}}_{\text{eff-NI}} \big\}=\boldsymbol{0}_{MN \times MN}$. Exploiting the above results, we have 
\begin{equation}
\label{eqn:Expxp}
\mathbb{E}\left\{\mathbf{e}_{p} \right\}=\boldsymbol{0}_{MN \times 1}.
\end{equation}
It can be readily verified that the $ \left( l+Mk \right) $th element of the vector $ \mathbf{e}_{p}$ is
\begin{align}\label{eq:e_p_ele}
\mathbf{e}_{p}\left(l+Mk\right)= \sum_{i=1}^{Q}\left(h_{i}-\hat{h}_{\text{NI},i}\right)\alpha_{i}\left(l+Mk\right)x_{p}\left(\left[l-l_{i} \right]_{M}, \left[k-k_{i} \right]_{N} \right).
\end{align}
The factor $\alpha_i\left(l+Mk\right)$ models the time-frequency domain ISI and ICI \cite{raviteja2018practical}, and is given as 
\begin{eqnarray}\label{eq:zeta_prop}
\alpha_i\left(l+Mk\right)= \left\{
 \begin{array}{@{}ll@{}}
  e^{-j2\pi \frac{k}{N}z^{k_{i}\left(\left[l-l_{i} \right]_{M} \right)}},  &\text{if} \ \  l < l_{i} \\
z^{k_{i}\left(\left[l-l_{i} \right]_{M} \right)}, &\text{if}\ \ l\geq l_{i} \\
0, & \text{otherwise}.
\end{array}\right.
\end{eqnarray}
This expression in \eqref{eq:e_p_ele}, along with the fact  that all  pilot symbols have equal power of $ \sigma_{p}^{2} $, is used to compute the variance of the $\left(l+Mk\right)$th element of the vector $ \mathbf{e}_{p} $ as
\begin{align}
\label{eqn:Sigma2}
\mbox{var}\left[\mathbf{e}_{p}\left(l+Mk\right)\right]=\sum_{i=1}^{Q}\mathbb{E}\left\{\left|h_{i}-\hat{h}_{\text{NI},i} \right|^{2} \right\}\sigma^{2}_{p} = \sigma_{p}^{2}\mathbb{E}\left\{\left\|\mathbf{h}-\hat{\mathbf{h}}_{\text{NI }} \right\|^{2} \right\} \stackrel{(a)}{=}\sigma_{p}^{2} B_{h,\: \text{NI }}.
\end{align}
Equality in (a) is due to \eqref{eqn:MSEin}.  Since the vectors $ \mathbf{e}_{p} $ and $ \tilde{\mathbf{w}} $ are statistically independent,  the mean and the variance of $ i $th element of  $ \tilde{\mathbf{w}}_{e} $ using (\ref{eqn:Covnoise}), (\ref{eqn:Expxp}) and (\ref{eqn:Sigma2}), can be computed as
\begin{eqnarray}
\label{eqn:expectwe1}
\mathbb{E}\left\{\tilde{\mathbf{w}}_{e}\left(i \right) \right\}&=&\mathbb{E}\left\{\mathbf{e}_{p}\left(i\right) \right\}+ \mathbb{E}\left\{\tilde{\mathbf{w}}\left(i\right) \right\}=0
\\\label{eqn:varwe1}
\mbox{var}\left[ \tilde{\mathbf{w}}_{e}\left(i \right) \right]&=&\mbox{var}\left[\mathbf{e}_{p}\left(i\right) \right]+\mbox{var}\left[\tilde{\mathbf{w}}\left(i\right) \right]= \sigma_{p}^{2} B_{h, \text{NI}}+\sigma_{w}^{2}. 
\end{eqnarray}
{Using  \eqref{eqn:expectwe1}}, and the channel estimate $\widehat{\mathbf{H}}_{\text{eff-NI}}$, the mean  $\mu^{(i)}_{a,b}$ of the interference-plus-noise term $\gamma_{a,b}$ can be
evaluated as follows
 \begin{align} \label{eqn:mean}
 \mu_{a,b}^{\left(i\right)}=\sum_{c \in \mathcal{I}\left(a\right), c\neq b}\sum_{j=1}^{S}p_{c,a}^{\left(i-1\right)}\left(\alpha_{j}\right)\alpha_{j}\widehat{\mathbf{H}}_{\text{eff-NI}}\left(a,c\right).
 \end{align}
Here we have used the property $\mathbb{E}\left\{\tilde{\mathbf{w}}_{e}\left(a\right)\right\}=0$ from \eqref{eqn:expectwe1}. The quantity $p_{c,a}(\alpha_j)$, which is calculated later in this section, denotes the pmf of the $j$th constellation symbol $\alpha_j$ between the observation node $a$ and the variable node $c$. The parameter $S$ denotes the constellation size. The variance of the noise-plus-interference term $\gamma_{a,b}$ for the $i$th iteration is next computed as follows
 \begin{align} \label{eqn:variance}
 &\left(\sigma_{a,b}^{\left(i\right)}\right)^{2}=\sum_{c \in \mathcal{I}\left(a\right), c\neq b}\left(\sum_{j=1}^{S}p_{c,a}^{\left(i-1\right)}\left(\alpha_{j}\right)\left|\alpha_{j}\right|^{2}\left|\widehat{\mathbf{H}}_{\text{eff-NI}}\left(a,c\right)\right|^{2} - \left|\mu_{a,b}^{\left(i\right)}\right|^{2}\right)  + \text{var}\left[\tilde{\mathbf{w}}_{e}\left(a\right)\right].
 \end{align}
The variance $\text{var}\left[\tilde{\mathbf{w}}_{e}\left(a\right)\right]=\sigma_{p}^{2}B_{h, \text{NI}}+ \sigma_{w}^{2}$ is derived in \eqref{eqn:varwe1}}. {We see that, unlike \cite{DBLP:journals/tvt/RavitejaPH19}, \cite[Eq. (31)]{raviteja2018interference},}    $\left(\sigma_{a,b}^{\left(i\right)}\right)^{2}$ is also a function of the MSE of the SP-NI channel estimator and the pilot power~$\sigma^2_p$.

 We now calculate the elements of the pmf vector $\mathbf{p}_{b,a}\in \mathbb{R}^{S \times 1} $, which is passed from the variable nodes $\mathbf{x}_d(b)$ to the observation node $\mathbf{y}_d(a)$, where $a\in\mathcal{J}(b)$. {The messages in terms of pmfs can be computed at each iteration using various methods e.g., belief propagation (BP) based algorithm \cite{SomDSCR11},  double loop methods \cite{yuille2001double} and damping method \cite{pretti2005message}. As explained in \cite{SomDSCR11}, the damping method has better convergence behavior. We, therefore, use the damping method from \cite{pretti2005message} to recursively update the pmf $ p_{b,a}^{\left(i\right)}$ in the $i$th iteration as follows \cite{raviteja2018interference} }
 \begin{equation}\label{eq:pmf}
 p_{b,a}^{\left(i\right)}\left(\alpha_{j}\right)=\Delta\tilde{p}^{\left(i\right)}_{b,a}\left(\alpha_{j}\right)+ \left(1-\Delta\right)p_{b,a}^{\left(i-1\right)}\left(\alpha_{j}\right).
 \end{equation} Here $ \Delta \in \left(0,1\right]  $ is the damping factor, which controls the convergence of the message passing algorithm, and $1\leq j\leq S$. The probability $ \tilde{p}^{\left(i\right)}_{b,a}\left(\alpha_{j}\right) $ is computed as \cite{raviteja2018interference}
  \begin{equation}\label{eq:pmf_1}
\tilde{p}^{\left(i\right)}_{b,a}\left(\alpha_{j}\right)\approx \prod_{  c \in \mathcal{J}\left(b\right), c\neq a}\frac{\beta^{\left(i\right)}_{c,b,j}}{\sum_{s=1}^{S}\beta^{\left(i\right)}_{c,b,s}},
  \end{equation}
  where $ \beta^{\left(i\right)}_{c,b,s}=\exp \left(\frac{-\left|\mathbf{y}_{d}\left(c\right)-\mu_{c,b}^{\left(i\right)}-\widehat{\mathbf{H}}_{\text{eff-NI}}\left(c,b\right)\alpha_{s}\right|}{\left(\sigma_{c,b}^{\left(i\right)}\right)^{2}}\right) $.
To terminate the message passing algorithm, the convergence indicator $ \zeta^{\left(i\right)} $ is given as~\cite{raviteja2018interference}
  \begin{equation}\label{eq:conver_I}
  \zeta^{\left(i\right)}=\frac{1}{MN}\sum_{b=1}^{MN}\mathbb{I}\left(\max_{a_{j}\in \mathbb{A}}p_{b}^{\left(i\right)}\left(\alpha_{j}\right)\geq 1 - \epsilon\right),
  \end{equation} 
  where $\mathbb{I}(\cdot)$ is the indicator function, $ \epsilon > 0 $ and 
  \begin{equation}
  p_{b}^{\left(i\right)}\left(\alpha_{j}\right)=\prod_{  c \in \mathcal{J}\left(b\right)}\frac{\beta^{\left(i\right)}_{c,b,j}}{\sum_{s=1}^{S}\beta^{\left(i\right)}_{c,b,s}}.
  \end{equation}
{If the convergence indicator $ \zeta^{\left(i\right)} > \zeta^{\left(i-1\right)} $, decision on the delay-Doppler data symbol in the SP-NI design for the  $ i $th iteration of the message passing algorithm is updated as 
 \begin{equation}
 \label{eqn:detecteddata}
 \hat{\mathbf{x}}_{d,\text{NI}}\left(b\right)=\argmax_{1\leq j \leq S} p_{b}^{\left(i\right)}\left(\alpha_{j}\right).
 \end{equation}}
Algorithm~\ref{algo:MP} summarizes the message passing  aided detection for the proposed SP-NI design. {Step~3 passes the mean $\mu_{a,b}^{\left(i\right)}$ and variance $(\sigma^{(i)}_{a,b})^2$ from the observation nodes to variable nodes. Step~4 updates the pmfs and pass them from the variable nodes to the observation nodes. This is followed by the data detection in the Step~5.}
	 \vspace{-0.5cm}
\begin{algorithm}[htbp]
\DontPrintSemicolon 
\scriptsize
\KwIn{ $ \mathbf{y}_{d} $, $ \widehat{\mathbf{H}}_{\text{eff-NI}} $}
\KwOut{$\hat{\mathbf{x}}_{d,\text{NI}}$}
\textbf{Initialization:} $\mathbf{p}^{(0)}_{b,a} = 1/S$, for $1\leq b\leq MN$ and $a\in\mathcal{J}(b)$, Counter $ i=1 $.\;
  \Repeat{Stopping criteria} {
    Pass the mean $\mu_{a,b}^{\left(i\right)}$ in \eqref{eqn:mean}  and variance $(\sigma^{(i)}_{a,b})^2$ in  \eqref{eqn:variance} from observation nodes $ \mathbf{y}_{d} $ to variable nodes $ \mathbf{x}_{d} (c) $ for $c\in\mathcal{I}(a)$ \;
At variable nodes $\mathbf{x}_d(b)$, update the pmf using \eqref{eq:pmf}, and pass it to the observation nodes $\mathbf{y}_d(a)$ for $a\in\mathcal{J}(b)$\;
{Update the decision on the transmitted symbols $ \mathbf{x}_{d} $, using \eqref{eqn:detecteddata}},  and obtain $ \hat{\mathbf{x}}_{d,\text{NI}} $ \;
$i=i+1$\;
  }
\Return:{\ $\hat{\mathbf{x}}_{d,\text{NI}}$}\;
\caption{{\sc }\small  Message passing  data detection algorithm for the proposed SP-NI design.}
\label{algo:MP}
\end{algorithm}
\vspace{-1.3cm}
 \subsection{Proposed superimposed pilot-iterative (SP-I) design}\label{SPI_Scheme}
The proposed SP-NI design, described in the previous section, estimates channel by treating data as  interference, which as shown later in Section-\ref{Results}, degrades its performance at high SNR values. To handle  this problem, we now propose the SP-I design, which iterates between channel estimation and message passing-aided data detection.  The SP-I design begins by taking the initial data estimate $\hat{\mathbf{x}}^{(0)}_{d}= \hat{\mathbf{x}}_{d,\text{NI}} $  from the  SP-NI design, and uses it along with the superimposed pilot vector $ \mathbf{x}_{p} $  to perform data-aided channel estimation, as discussed next.
	 \vspace{-0.4cm}
\subsubsection{Channel estimation in SP-I design}
Let $\widehat{\boldsymbol{\Omega}}^{(0)}_{d}$ be the initial estimate of the data matrix ${\boldsymbol{\Omega}}_{d}$. It is obtained from SP-NI design  by substituting the data estimate $\mathbf{x}_{d}=\hat{\mathbf{x}}^{(0)}_d=\hat{\mathbf{x}}_{d,\text{NI}}$  in \eqref{eqn:data}~as
\begin{align}
\label{eqn:omegadata1}
\widehat{\boldsymbol{\Omega}}^{(0)}_{d}&= \left[\boldsymbol{\Gamma}_{1}\hat{\mathbf{x}}^{(0)}_{d} \;  \; \boldsymbol{\Gamma}_{2}\hat{\mathbf{x}}^{(0)}_{d} \;  \; \cdots \;  \; \boldsymbol{\Gamma}_{Q}\hat{\mathbf{x}}^{(0)}_{d} \right].
\end{align}
The received signal  vector in (\ref{eqn:inopy}) can now be re-expressed as 
 \begin{eqnarray}\label{eq:DA_model}
 \mathbf{y}=\big(\boldsymbol{\Omega}_{p} + \widehat{\boldsymbol{\Omega}}^{(0)}_{d}\big)\mathbf{h}+\big(\boldsymbol{\Omega}_{d} - \widehat{\boldsymbol{\Omega}}^{(0)}_{d}\big)\mathbf{h}+\tilde{\mathbf{w}}=  \boldsymbol{\Omega}^{(0)}_{\mathbf{x}_{p}\hat{\mathbf{x}}_{d}}\mathbf{h}+\boldsymbol{\xi}^{(0)}_{\tilde{\mathbf{w}}}.
 \end{eqnarray}
The data-aided pilot matrix $\boldsymbol{\Omega}^{(0)}_{\mathbf{x}_{p}\hat{\mathbf{x}}_{d}}$ and the noise-plus-interference vector $\boldsymbol{\xi}^{(0)}_{\tilde{\mathbf{w}}}$ are defined as
\begin{eqnarray}\label{eqn:xi1}
\boldsymbol{\Omega}^{(0)}_{\mathbf{x}_{p}\hat{\mathbf{x}}_{d}}\overset{\Delta}{=}\boldsymbol{\Omega}_{p} + \widehat{\boldsymbol{\Omega}}^{(0)}_{d} \ \ \text{and} \ \ 
\boldsymbol{\xi}^{(0)}_{\tilde{\mathbf{w}}}\overset{\Delta}{=} \boldsymbol{\Xi }^{(0)}_{\mathbf{x}_{d}}\mathbf{h}+\tilde{\mathbf{w}},
\end{eqnarray}
with the estimation error matrix 
$\boldsymbol{\Xi }^{(0)}_{\mathbf{x}_{d}}\overset{\Delta}{=}\boldsymbol{\Omega}_{d} -\widehat{\boldsymbol{\Omega}}^{(0)}_{d}$.
Using the above initial data estimate, the SP-I design performs data-aided estimation of the channel vector $ \mathbf{h} $ in \eqref{eq:DA_model} in the $n$th iteration~as 
\begin{eqnarray}
\label{eqn:hathdata44}
\hat{\mathbf{h}}^{(n)}=\left(\big(\boldsymbol{\Omega}^{(n-1)}_{\mathbf{x}_{p}\hat{\mathbf{x}}_{d}}\big)^H\big(\mathbf{C}^{(n-1)}_{\boldsymbol{\xi}_{\tilde{\mathbf{w}}}}\big)^{-1} \boldsymbol{\Omega}^{(n-1)}_{\mathbf{x}_{p}\hat{\mathbf{x}}_{d}}+ \mathbf{C}^{-1}_{\mathbf{h}}\right)^{-1}\big(\boldsymbol{\Omega}^{(n-1)}_{\mathbf{x}_{p}\hat{\mathbf{x}}_{d}}\big)^H\big(\mathbf{C}^{(n-1)}_{\boldsymbol{\xi}_{\tilde{\mathbf{w}}}}\big)^{-1}\mathbf{y}.
\end{eqnarray}
Since $ \mathbb{E}\left\{\mathbf{h} \right\}=\boldsymbol{0}_{Q\times 1} $, we have from \eqref{eqn:Covnoise} that $\mathbb{E}\left\{\boldsymbol{\xi}^{(n)}_{\tilde{\mathbf{w}}} \right\}= \boldsymbol{0}_{MN \times 1}$. The covariance matrix  $\mathbf{C}^{(n)}_{\boldsymbol{\xi}_{\tilde{\mathbf{w}}}}$ of the vector $\boldsymbol{\xi}^{(n)}_{\tilde{\mathbf{w}}}$ in the $n$th iteration is computed in the next lemma, which is proved in Appendix-\ref{Appendix_C}.
	 \vspace{-0.7cm}
\begin{lemma}\label{Lemma3}
The covariance matrix $\mathbf{C}^{(n)}_{\boldsymbol{\xi}_{\tilde{\mathbf{w}}}}=\mathbb{E}\big\{ \boldsymbol{\xi}^{(n)}_{\tilde{\mathbf{w}}}\big(\boldsymbol{\xi}_{\tilde{\mathbf{w}}}^{(n)}\big)^H\big\}$ of the vector $ \boldsymbol{\xi}^{(n)}_{\tilde{\mathbf{w}}} $ is 
\begin{eqnarray}
\label{eqn:Covxiw1}
\mathbf{C}^{(n)}_{\boldsymbol{\xi}_{\tilde{\mathbf{w}}}}
= 2\left(\sum_{i=1}^{Q}\sigma_{h_{i}}^{2} \right)\sigma_{d}^{2}\mathbf{I}_{MN}+\sigma_{w}^{2}\mathbf{I}_{MN}.
\end{eqnarray}
\end{lemma}
The MSE of the data-aided channel estimator is \cite{kay1993fundamentals}:
 \begin{equation}\label{eq:MSE_SPI}
 B^{(n)}_{h}=\mbox{Tr}\left(\boldsymbol{\Sigma}^{(n)}_{\mathbf{h}} \right),
 \end{equation}
where $\boldsymbol{\Sigma}^{(n)}_{\mathbf{h}}
= \left(\big(\boldsymbol{\Omega}^{(n-1)}_{\mathbf{x}_{p}\hat{\mathbf{x}}_{d}}\big)^H\big(\mathbf{C}^{(n-1)}_{\boldsymbol{\xi}_{\tilde{\mathbf{w}}}}\big)^{-1}\boldsymbol{\Omega}^{(n-1)}_{\mathbf{x}_{p}\hat{\mathbf{x}}_{d}}+ \mathbf{C}^{-1}_{\mathbf{h}} \right)^{-1}
$ is the covariance matrix of the error $\big(\mathbf{h}-\hat{\mathbf{h}}^{(n)}\big)$.
Let $\widehat{\mathbf{H}}^{(n)}_{\text{eff}}$ be the estimate of the effective channel matrix corresponding to the data-aided channel estimate $ \hat{\mathbf{h}} ^{(n)}$ in the $n$th iteration. It is obtained by substituting  $ \hat{\mathbf{h}}^{(n)} $ in \eqref{eqn:heff}:
\begin{equation}\label{eqn:H_eff11}
\widehat{\mathbf{H}}^{(n)}_{\text{eff}}=\mathbf{B}_{\text{rx}}\left(\sum_{i=1}^{Q}\hat{h}^{(n)}_i\boldsymbol{\Theta}_{i}\right)\mathbf{B}_{\text{tx}}.
\end{equation}
\subsubsection{Message passing-aided data detection in SP-I design}
\label{Section_MP2}
	 \vspace{-0.3cm}
We now use the message passing algorithm to detect data by using data-aided channel estimate. The receive signal $ \tilde{\mathbf{y}}^{(n)}_{d}\in \mathbb{C}^{MN \times 1} $ used for  detecting data in the $n$th  SP-I iteration  is obtained from the received vector in (\ref{eqn:rxvectr12})  as
\begin{eqnarray}
\tilde{\mathbf{y}}^{(n)}_{d}=\mathbf{y}-\widehat{\mathbf{H}}^{(n)}_{\text{eff}}\mathbf{x}_{p} =\mathbf{H}_{\text{eff}}\mathbf{x}_{d}+\tilde{\mathbf{e}}^{(n)}_{p}+\tilde{\mathbf{w}} = \mathbf{H}_{\text{eff}}\mathbf{x}_{d}+\tilde{\mathbf{w}}^{(n)}_{\tilde{e}},
\end{eqnarray}
where the error vector $\tilde{\mathbf{e}}^{(n)}_{p}=\big(\mathbf{H}_{\text{eff}}-\widehat{\mathbf{H}}^{(n)}_{\text{eff}} \big)\mathbf{x}_{p}$ and the noise-plus-estimation error vector $\tilde{\mathbf{w}}^{(n)}_{\tilde{e}}=\tilde{\mathbf{e}}^{(n)}_{p}+\tilde{\mathbf{w}}$. \textit{We see that statistics  $\mathbf{e}_p^{(n)}$, in each iteration, is also a function of data estimate. This is because the channel estimate, as shown in \eqref{eqn:hathdata44}, is a function of data estimate. The first- and second order characteristics of $\mathbf{e}_p^{(n)}$ now also need the first- and second-order statistics of the data estimate.} The mean and variance of the $ j $th element of $ \tilde{\mathbf{e}}^{(n)}_{p} $, similar to the derivation of (\ref{eqn:Expxp}) and (\ref{eqn:Sigma2}),  can be obtained as $\mathbb{E}\left\{ \tilde{\mathbf{e}}^{(n)}_{p}\left(j\right)\right\}=0$ and $\mbox{Var}\left[\tilde{\mathbf{e}}^{(n)}_{p}\left(j\right) \right]=\sigma_{p}^{2}B^{(n)}_{h}$, respectively. Furthermore, since $ \tilde{\mathbf{e}}^{(n)}_{p} $ and $ \tilde{\mathbf{w}} $ are statistically independent random vectors, we get 
\begin{equation}\label{eq:mu_sigma}
\mathbb{E}\left\{\tilde{\mathbf{w}}^{(n)}_{\tilde{\mathbf{e}}}\left(j\right) \right\}=0\  \text{and } \mbox{Var}\left[ \tilde{\mathbf{w}}^{(n)}_{\tilde{\mathbf{e}}}\left(j\right)\right]=\sigma_{p}^{2}B^{(n)}_{h}+\sigma_{w}^{2}.
\end{equation}
{For detecting data  with the channel estimate $\widehat{\mathbf{H}}^{(n)}_{\text{eff}}$, the mean $\mu_{a,b}^{\left(i\right)}$ in the $i$th iteration of the message passing algorithm, similar to \eqref{eqn:mean}, is calculated as follows
 \begin{align} \label{eqn:mean1}
 \mu_{a,b}^{\left(i\right)}&=\sum_{c \in \mathcal{I}\left(a\right), c\neq b}\sum_{j=1}^{S}p_{c,a}^{\left(i-1\right)}\left(\alpha_{j}\right)\alpha_{j}\widehat{\mathbf{H}}^{(n)}_{\text{eff}}\left(a,c\right) +\mathbb{E}\left\{\tilde{\mathbf{w}}^{(n)}_{\tilde{e}}\left(a\right)\right\}
 \end{align}
Here, as shown in \eqref{eq:mu_sigma},  $\mathbb{E}\left\{\tilde{\mathbf{w}}^{(n)}_{\tilde{e}}\left(a\right)\right\}=0$. The variance $\left(\sigma_{a,b}^{\left(i\right)}\right)^{2}$ for the $i$th iteration of the message passing design,  using \eqref{eq:mu_sigma} and the estimate $\widehat{\mathbf{H}}^{(n)}_{\text{eff}}$, is given as follows
 \begin{align} \label{eqn:variance1}
 &\left(\sigma_{a,b}^{\left(i\right)}\right)^{2}=\sum_{c \in \mathcal{I}\left(a\right), c\neq b}\left(\sum_{j=1}^{S}p_{c,a}^{\left(i-1\right)}\left(\alpha_{j}\right)\left|\alpha_{j}\right|^{2}\left|\widehat{\mathbf{H}}^{(n)}_{\text{eff}}\left(a,c\right)\right|^{2} - \left|\mu_{a,b}^{\left(i\right)}\right|^{2}\right)  + \sigma_{p}^{2}B^{(n)}_{h}+ \sigma_{w}^{2}.
 \end{align}}{The  pmf $p_{c,a}^{(i)}$ is calculated from \eqref{eq:pmf} and \eqref{eq:pmf_1} by  using the channel estimate $\widehat{\mathbf{H}}^{(n)}_{\text{eff}}$, and the above computed mean $ \mu_{a,b}^{\left(i\right)}$ and variance $\left(\sigma_{a,b}^{\left(i\right)}\right)^{2}$. We once again observe that, unlike \cite{raviteja2018interference,DBLP:journals/tvt/RavitejaPH19}, the variance $\left(\sigma_{a,b}^{\left(i\right)}\right)^{2}$ is also a function of the MSE $B^{(n)}_h$ of the proposed SP-I channel estimator and the pilot power $\sigma^2_p$.} The computed mean and variance, along with the vector $\tilde{\mathbf{y}}^{(n)}_d$ and estimate $\widehat{\mathbf{H}}^{(n)}_{\text{eff}}$, are fed to Algorithm~\ref{algo:MP} for calculating the data estimate $\hat{\mathbf{x}}^{(n)}_d$. 
Algorithm~\ref{algo:Summary} summarizes the proposed SP-I design. {This algorithm is initialized with data detected by the SP-NI design.  The parameters in Steps~$3-4$ are computed using data detected in $(n-1)$th iterations. Step~$5$ describes the iterative data-aided MMSE channel estimation in $n$th iteration. The next step feeds data-aided channel estimate $\widehat{\mathbf{H}}^{(n)}_{\text{eff}}$, along with the vector $\tilde{\mathbf{y}}^{(n)}_d$, to  Algorithm-\ref{algo:MP} for calculating $\hat{\mathbf{x}}^{(n)}_d$.} 
\begin{algorithm}[htbp]
\DontPrintSemicolon 
\scriptsize
\KwIn{ Observation vector $\mathbf{y}$, pilot matrix $\boldsymbol{\Omega}_{p}$, $\mathbb{E}\{\mathbf{h}\}$, $\mathbf{C}_{\mathbf{h}}=\mathbb{E}\{\mathbf{h}\mathbf{h}^{H}\}$ and the initial data estimate $\hat{\mathbf{x}}^{(0)}_d=\hat{\mathbf{x}}_{d,\text{NI}}$ from Algorithm-\ref{algo:MP}.}
\KwOut{Channel estimate $\hat{\mathbf{h}}$ and data estimate $\hat{\mathbf{x}}_d$}
\textbf{Initialize:} Counter $n=1$\;
  \Repeat{Stopping criteria} {
  Compute $\widehat{\boldsymbol{\Omega}}_{d}^{(n-1)}$ as given in \eqref{eqn:omegadata1}\;
 Compute $\boldsymbol{\Omega}_{\mathbf{x}_{p}\hat{\mathbf{x}}_{d}}^{(n-1)}$  and $ \mathbf{C}^{(n-1)}_{\boldsymbol{\xi}_{\tilde{\mathbf{w}}}} $ as derived in \eqref{eqn:xi1} and \eqref{eqn:Covxiw1}, respectively\;
Compute $ \hat{\mathbf{h}}^{(n)}$ and $\widehat{\mathbf{H}}^{(n)}_{\text{eff}}$ as derived in~\eqref{eqn:hathdata44} and \eqref{eqn:H_eff11}, respectively\;
Feed the estimate $\widehat{\mathbf{H}}^{(n)}_{\text{eff}}$ and the vector  $\tilde{\mathbf{y}}^{(n)}_d$ to the message passing receiver  Algorithm~\ref{algo:MP} to calculate the data estimate $\hat{\mathbf{x}}^{(n)}_d$. Note that the messages $\mu^{(i)}_{a,b}$ and $\left(\sigma^{(i)}_{a,b}\right)^2$  for the $i$th iteration of the message passing algorithm are computed from \eqref{eqn:mean1}  and \eqref{eqn:variance1}, respectively. The corresponding pmf $p_{c,a}^{(i)}$ is evaluated by using $\widehat{\mathbf{H}}^{(n)}_{\text{eff}}$ and the messages $ \mu_{a,b}^{\left(i\right)}$ and $\left(\sigma_{a,b}^{\left(i\right)}\right)^{2}$ in \eqref{eq:pmf} and \eqref{eq:pmf_1}. \;
$n=n+1$\;
  }
\Return:{\ $\hat{\mathbf{x}}_d$}\;
\caption{{\sc } \small Iterative channel and data detection for the SP-I design .}
\label{algo:Summary}
\end{algorithm}
\vspace{-0.7cm}
\section{Optimal power allocation between data and pilot symbols} \label{sec:optimal_power}
The proposed  designs  superimpose  pilots on to  data symbols. {Given the total power constraint,  we now optimally allocate power between data and pilots  to maximize SINR, which will consequently minimize BER and maximize SE \cite{SinghMJVH20}.} We, therefore, first derive the SINR using the SP-NI channel estimate $\hat{\mathbf{h}}_{\text{NI }}$, which  is then maximized to calculate optimal pilot power $\sigma^2_{p,\text{opt}}$ and  data power  $\sigma^2_{p,\text{opt}}=1-\sigma^2_{d,\text{opt}}$. We will show that this optimal power allocation also maximizes the SE and minimizes  the BER of SP-I design also. {The SINR, as shown next, is a function of  matrices $\mathbf{\Omega}_d$ and $\mathbf{\Omega}_p$, and therefore, also a function of $\boldsymbol{\Gamma}_{i}=\mathbf{B}_{\text{rx}}\boldsymbol{\Theta}_{i}\mathbf{B}_{\text{tx}}$, in addition to data and pilot symbols. The noise-plus-interference vector $\check{\mathbf{w}}$ is also a function of $\mathbf{\Gamma}_i$ matrices through $\mathbf{\Omega}_d$ and $\mathbf{\Omega}_p$. By using ideas from random matrix theory, and inequalities/ideas from linear algebra, we now derive the SINR lower bound.  Since the SINR is also a function of channel estimate, it requires  its MSE.} We calculate the effective SINR by using the channel estimate $\hat{\mathbf{h}}_{\text{NI}}$ derived in \eqref{eqn:hathinitial33}.  The observation vector $\mathbf{y}_d$ in \eqref{eqn:yd1}, using \eqref{eqn:inopy},  can be rewritten~as follows
\begin{eqnarray}
\label{eqn:effecdata}
\mathbf{y}_{d}&=& \mathbf{y}-\boldsymbol{\Omega}_{p}\hat{\mathbf{h}}_{\text{NI }} = \boldsymbol{\Omega}_{d}\mathbf{h}+ \left(\mathbf{h} -\hat{\mathbf{h}}_{\text{NI }} \right)\boldsymbol{\Omega}_{p}+\tilde{\mathbf{w}} \nonumber \\
&=& \boldsymbol{\Omega}_{d}\hat{\mathbf{h}}_{\text{NI }}+\left(\boldsymbol{\Omega}_{d}+\boldsymbol{\Omega}_{p}\right)\tilde{\mathbf{h}}+ \tilde{\mathbf{w}} = \boldsymbol{\Omega}_{d}\hat{\mathbf{h}}_{\text{NI }}+\check{\mathbf{w}}.
\end{eqnarray}
The error vector $\tilde{\mathbf{h}}$ and the noise-plus-interference vector $\check{\mathbf{w}}$ are defined as
\begin{eqnarray}
\label{eqn:esterror}
\tilde{\mathbf{h}}=\mathbf{h}-\hat{\mathbf{h}}_{\text{NI }} \text{ and }
\check{\mathbf{w}}=\left(\boldsymbol{\Omega}_{d}+\boldsymbol{\Omega}_{p}\right)\tilde{\mathbf{h}}+\tilde{\mathbf{w}}.
\end{eqnarray}
{The received symbol at delay-Doppler location $(l,k)$, using \eqref{eqn:effecdata}, can be expressed as \cite{raviteja2018practical}
\begin{align}\label{eq:y_kl}
\nonumber y_{d}\left[l,k\right]& = \sum_{i=1}^{Q}\hat{h}_{\text{NI},i}^{\alpha} x_{d}\left[\left(l-l_{i}\right)_{M}, \left(k-k_{i}\right)_{N}\right]+ \sum_{i=1}^{Q} \tilde{h}_{\text{NI},i}^{\alpha}\Big(x_{d}\left[\left(l-l_{i}\right)_{M}, \left(k-k_{i}\right)_{N}\right]\\
&+x_{p}\left[\left(l-l_{i}\right)_{M}, \left(k-k_{i}\right)_{N}\right]\Big)+w\left[l,k\right],
\end{align}
where $ \hat{h}_{\text{NI},i}^{\alpha}= \hat{h}_{\text{NI},i}\alpha_{i}\left(l,k\right)$ and  $ \tilde{h}_{\text{NI},i}^{\alpha}=\left(h_{i}-\hat{h}_{\text{NI},i}\right)\alpha_{i}\left(l,k\right) $. Equation \eqref{eq:y_kl} can be re-written as 
\begin{align}
y_{d}\left[l,k\right]= \tilde{\mathbf{x}}_{d}^{T}\hat{\mathbf{h}}_{\text{NI}}^{\alpha}+\left(\tilde{\mathbf{x}}_{d}+\tilde{\mathbf{x}}_{p}\right)^{T}\tilde{\mathbf{h}}_{\text{NI}}^{\alpha}+w\left[l,k\right] = \tilde{\mathbf{x}}_{d}^{T}\hat{\mathbf{h}}_{\text{NI}}^{\alpha}+ v\left[l,k\right].
\end{align}}{The noise-plus-interference term $ v\left[l,k\right]=\left(\tilde{\mathbf{x}}_{d}+\tilde{\mathbf{x}}_{p}\right)^{T}\tilde{\mathbf{h}}_{\text{NI}}^{\alpha}+w\left[l,k\right], $s and  $ \tilde{\mathbf{x}}_{d} \in \mathbb{C}^{Q\times 1}$ and $ \tilde{\mathbf{x}}_{p}\in \mathbb{C}^{Q \times 1} $ are data and pilot vectors. Their $i$th element is given as $x_{d}\left[\left(l-l_{i}\right)_{M}, \left(k-k_{i}\right)_{N}\right]$ and $x_{p}\left[\left(l-l_{i}\right)_{M}, \left(k-k_{i}\right)_{N}\right]$, respectively. The scalers $ \hat{h}_{\text{NI},i}^{\alpha} $ and $ \tilde{h}_{\text{NI},i}^{\alpha} $ denote the $ i $th element of $ \hat{\mathbf{h}}_{\text{NI}}^{\alpha} $ and $ \tilde{\mathbf{h}}_{\text{NI}}^{\alpha}  $, respectively. The SINR for $(l,k)$th delay-Doppler domain symbol is now given as 
\begin{equation}
\label{eq:SINR}
\text{SINR}_{l,k}= \dfrac{\mathbb{E}\Big\{\big|\tilde{\mathbf{x}}_{d}^{T}\hat{\mathbf{h}}_{\text{NI}}^{\alpha} \big|^{2}\Big\}}{\mathbb{E}\Big\{\big|v\left[l,k\right] \big|^{2}\Big\}}.
\end{equation} }{Its numerator and denominator are a function of the channel estimate and estimation error, respectively. For a linear MMSE estimator, the estimation error is orthogonal to the observations \cite{kay1993fundamentals}. The noise is also independent of data symbols. These properties imply that the numerator and denominator of  \eqref{eq:SINR} are independent. The numerator of \eqref{eq:SINR} is simplified next.
\begin{align}
\mathbb{E}\Big\{\big|\tilde{\mathbf{x}}_{d}^{T}\hat{\mathbf{h}}_{\text{NI}}^{\alpha} \big|^{2}\Big\}&=\mathbb{E}\Big\{\big(\hat{\mathbf{h}}_{\text{NI}}^{\alpha}\big)^{H}\mathbb{E}\left\{\tilde{\mathbf{x}}_{d}^{*}\tilde{\mathbf{x}}_{d}^{T}\right\} \hat{\mathbf{h}}_{\text{NI}}^{\alpha}\Big\} \nonumber \stackrel{\left(b\right)}{=} \sigma_{d}^{2} \mathbb{E}\Big\{\big\|\hat{\mathbf{h}}_{\text{NI}} \big\|^{2}\Big\}.
\end{align}
Equality $ \left(b\right) $ is because $ \mathbb{E}\left\{\tilde{\mathbf{x}}_{d}^{*}\tilde{\mathbf{x}}_{d}^{T}\right\}=\sigma_{d}^{2}\mathbf{I}_{Q} $ and $ \alpha_{i}\left(l,k\right) $ is a phase factor. The quantity $\mathbb{E}\Big\{\big\|\hat{\mathbf{h}}_{\text{NI}} \big\|^{2}\Big\}=\mbox{Tr}\Big(\mathbb{E}\Big\{\hat{\mathbf{h}}_{\text{NI }}\big(\hat{\mathbf{h}}_{\text{NI }}\big)^{H}\Big\}\Big)$ can be calculated using \eqref{eqn:esterror}~as follows 
 \begin{align}
 \label{eqn:trhhathhatHe}
\text{Tr}\Big(\mathbb{E}\big\{\hat{\mathbf{h}}_{\text{NI}}\big(\hat{\mathbf{h}}_{\text{NI}}\big)^{H}\big\}\Big)= \sigma_{h}^{2}-B_{h,\text{NI}}.
 \end{align}}{Here $\sigma_{h}^{2}=\mbox{Tr}(\mathbf{C}_{\mathbf{h}})=\sum_{i=1}^{Q} \sigma_{h_{i}}^{2} $ and $ B_{h, \text{NI}}=\mbox{Tr}\left(\boldsymbol{\Sigma}_{\text{NI }}\right) $ is derived in \eqref{eqn:MSEin}. 
	 The above equation implies that the numerator of SINR expression in \eqref{eq:SINR} is  
\begin{align}
\label{eq:sigpow2}
\mathbb{E}\big\{|\tilde{\mathbf{x}}_{d}^{T}\hat{\mathbf{h}}_{\text{NI}}^{\alpha} |^{2}\big\}= \sigma_{d}^{2} \left(\sigma_{h}^{2}-B_{h,\text{NI}}\right).
\end{align}
We next simplify the denominator of SINR expression in \eqref{eq:SINR} as follows
\begin{align}
\label{eq:Intrferenec}
\mathbb{E}\left\{\left|v\left[l,k\right] \right|^{2}\right\}&=\mathbb{E}\left\{\left(\tilde{\mathbf{h}}_{\text{NI}}^{\alpha}\right)^{H}\left(\mathbb{E}\left\{\tilde{\mathbf{x}}_{d}^{*}\tilde{\mathbf{x}}_{d}^{T}\right\}+\tilde{\mathbf{x}}_{p}^{*}\tilde{\mathbf{x}}_{p}^{T}\right)\tilde{\mathbf{h}}_{\text{NI}}^{\alpha}\right\}+\mathbb{E}\left\{\left| w\left[l,k\right]\right|^{2}\right\} \nonumber \\
&=\sigma_{d}^{2}\mathbb{E}\left\{\left\|\tilde{\mathbf{h}}_{\text{NI}} \right\|^{2}\right\}+\text{Tr}\left(\mathbb{E}\left\{\tilde{\mathbf{h}}_{\text{NI}}^{\alpha}\left(\tilde{\mathbf{h}}_{\text{NI}}^{\alpha}\right)^{H}\right\}\tilde{\mathbf{x}}_{p}^{*}\tilde{\mathbf{x}}_{p}^{T}\right)+\sigma_{w}^{2}.
\end{align}}{We now simplify the second term of the above expression as  
\begin{align}
\label{eq:interfenece2}
\text{Tr}\left(\mathbb{E}\left\{\tilde{\mathbf{h}}_{\text{NI}}^{\alpha}\left(\tilde{\mathbf{h}}_{\text{NI}}^{\alpha}\right)^{H}\right\}\tilde{\mathbf{x}}_{p}^{*}\tilde{\mathbf{x}}_{p}^{T}\right)\leq \text{Tr}\left(\mathbb{E}\left\{\tilde{\mathbf{h}}_{\text{NI}}^{\alpha}\left(\tilde{\mathbf{h}}_{\text{NI}}^{\alpha}\right)^{H}\right\}\right)\text{Tr}\left(\tilde{\mathbf{x}}_{p}^{*}\tilde{\mathbf{x}}_{p}^{T}\right)=Q\sigma^2_pB_{h,\text{NI}}. 
\end{align}
 The above inequality is because for two positive semi-definite matrices $ \mathbf{A}\in \mathbb{C}^{m\times m} $ and $ \mathbf{B}\in \mathbb{C}^{m\times m} $, the following holds i.e., $ \text{Tr}\left(AB\right) \leq \text{Tr}\left(A\right)\text{Tr}\left(B\right)$ \cite{ulukok2010some}.  Using \eqref{eq:interfenece2}, we simplify $ \mathbb{E}\left\{\left|v\left[l,k\right] \right|^{2}\right\}$ as
 \begin{align}
 \label{eq:noisepow}
 \mathbb{E}\left\{\left|v\left[l,k\right] \right|^{2}\right\}\leq \sigma_{d}^{2}B_{h, \text{NI}}+Q\sigma_{p}^{2}B_{h,\text{NI}}+\sigma_{w}^{2}.
 \end{align}}
The SINR in \eqref{eq:SINR} is computed on substitution of  \eqref{eq:sigpow2} and \eqref{eq:noisepow} as 
 \begin{align}\label{eqn:snr1}
 \text{SINR}\geq \dfrac{\sigma_{d}^{2} \left(\sigma_{h}^{2}-B_{h,\text{NI}}\right)}{\sigma_{d}^{2}B_{h, \text{NI}}+Q\sigma_{p}^{2}B_{h,\text{NI}}+\sigma_{w}^{2}}
 \end{align}
We have dropped the subscript $(l,k)$ as the SINR expression is independent of these indices. To evaluate the SINR expression in  \eqref{eqn:snr1}, we now calculate the MSE $ B_{h, \text{NI}} $ of the SP-aided channel estimation. It follows from  \eqref{eqn:MSEin} that $ B_{h, \text{NI}}= \mbox{Tr}\left[ \boldsymbol{\Sigma}_{\text{NI }}\right]$. We next propose the following lemma.
{\begin{lemma}\label{Lemma4}
	The MSE $ B_{h, \text{NI}} $ is lower bounded as 
	\begin{equation}
	\label{eqn:mse31}
	B_{h, \text{NI}} \geq \dfrac{Q^{2}}{\frac{QMN\sigma_{p}^{2}}{\sigma_{h}^{2}\sigma_{d}^{2}+\sigma_{w}^{2}}+ \tilde{\sigma}_{h}^{2}}.
	\end{equation}
\end{lemma}
\begin{IEEEproof}
	Refer to Appendix-\ref{Appendix_D}.
\end{IEEEproof}}

By substituting \eqref{eqn:mse31} in \eqref{eqn:snr1}, and by substituting $ \sigma_{d}^{2} = 1-\sigma_{p}^{2} $, the effective SINR is
 \begin{equation}
 \mbox{SINR}_{\text{eff}} \geq \dfrac{\sigma_{p}^{4}N_{1}+\sigma_{p}^{2}N_{2}+N_{3}}{\sigma_{p}^{4}D_{1}+\sigma_{p}^{2}D_{2}+D_{3}}.
 \end{equation}
Here 
$  N_{1}=\sigma_{h}^{2}QMN-\sigma_{h}^{2}\tilde{\sigma}_{h}^{2}+\sigma_{h}^{2}Q^{2}, 
 N_{2}=\sigma_{h}^{2}QMN-2\sigma_{h}^{2}\tilde{\sigma}_{h}^{2}+\sigma_{h}^{2}Q^{2}-\tilde{\sigma}_{h}^{2}\sigma_{w}^{2} +Q\sigma_{h}^{2}+Q^{2}\sigma_{w}^{2}$ and $
 N_{3}= \sigma_{h}^{2}\tilde{\sigma}_{h}^{2}+\tilde{\sigma}_{h}^{2}\sigma_{w}^{2}-Q^{2}\sigma_{h}^{2}-Q^{2}\sigma_{w}^{2}$. And, 
$ D_{1}=\sigma_{h}^{2}Q^{2}-\sigma_{h}^{2}Q^{3}, 
 D_{2}= \sigma_{h}^{2}Q^{3}+\sigma_{w}^{2}Q^{3}-2\sigma_{h}^{2}Q^{2}-\sigma_{w}^{2}Q^{2} 
+\sigma_{w}^{2}QMN-\sigma_{h}^{2}\tilde{\sigma}_{h}^{2}\sigma_{w}^{2}$ and $ 
 D_{3}= Q^{2}\sigma_{h}^{2}+Q^{2}\sigma_{w}^{2}+\sigma_{h}^{2}\tilde{\sigma}_{h}^{2}\sigma_{w}^{2}+\tilde{\sigma}_{h}^{2}\sigma_{w}^{4}.
 $
To calculate the optimal pilot power, we take the derivative of the lower bound on $\mbox{SINR}_{\text{eff}}$, and equate it to zero:
\begin{align}\label{eq:SINR_diff}
 \dfrac{\partial \mbox{SNR}_{\text{eff}}}{\partial \sigma_{p}^{2}}&=\sigma_{p}^{4}\left(D_{2}N_{1}-D_{1}N_{2}\right)+ \sigma_{p}^{2}\left(2D_{3}N_{1}-2D_{1}N_{3}\right)+ \left(D_{3}N_{2}-D_{2}N_{3}\right)=0.
\end{align}
After solving the above expression, the optimal pilot power is obtained as $ \sigma_{p,\text{opt}}^{2}=\Big|\dfrac{-b+\sqrt{b^{2}-4ac}}{2a} \Big|$,  where $a=D_{2}N_{1}-D_{1}N_{2}$, $b=2D_{3}N_{1}-2D_{1}N_{3}$ and $c=D_{3}N_{2}-D_{2}N_{3}$.  Using the constraint on total power per symbol, i.e., $\sigma^2_p+\sigma^2_d=1$, the optimal data power is obtained as  $\sigma^{2}_{d,\text{opt}}=1-\sigma^{2}_{p,\text{opt}}$.}
\vspace{-0.7cm}
\begin{remark}
	We note that the SINR is commonly calculated in the SP literature by averaging over estimated channel \cite{he2007superimposed}. This is because a major aim of deriving SINR is to optimally allocate power between data and pilot symbols to minimize BER. Since pilots are used to estimate channel, their power thus cannot depend on  instantaneous channel \cite{he2007superimposed}. {The SINR determines the allocated pilot power and consequently, cannot be a function of the instantaneous channel.}
\end{remark}
\vspace{-0.6cm}
\section{Computational Complexity Analysis}
\begin{footnotesize}
\begin{table*}
    \centering
    \caption{{Computational complexities of the proposed and the existing EP schemes.}}\label{tab:a}
    \begin{tabularx}{\linewidth}{|L|}
    \hline
 \textbf{Proposed SP-NI Scheme} \\ [0.5ex]
 \hline
\end{tabularx}
\begin{tabularx}{\linewidth}{|c|L|L|L|}
    \hline
 Operation & Complex Multiplications &Complex Additions & Total number of operations  \\ [0.5ex]
 \hline
$\mathbf{A}=\mathbf{\Omega}_{p}^{H}\mathbf{C}_{\tilde{\mathbf{w}}_d}^{-1}\mathbf{\Omega}_{p}+\mathbf{C}_{h}^{-1}$
& $Q^2MN+Q^2+Q$&$Q^2(MN-1)+Q^2$ & $2Q^2MN+Q^2+Q$\\
 \hline
$\mathbf{A}^{-1}$ &-&-&$\mathcal{O}(Q^3)$\\
 \hline
  $\mathbf{b}=\mathbf{\Omega}_{p}^{H}\mathbf{C}_{\tilde{\mathbf{w}}_d}^{-1}\mathbf{y}$
& $Q(MN+1)$& $Q(MN-1)$ & $2QMN$\\
 \hline
 $\mathbf{A}^{-1}\mathbf{b}$
& $Q^2$& $Q(Q-1)$ & $2Q^2-Q$\\
 \hline
  Data Detection using MP
& -& - & $\mathcal{O}(N_IMNQS)$ \cite{raviteja2018interference}\\
 \hline
\end{tabularx}
    \begin{tabularx}{\linewidth}{|L|}
    \hline
\textbf{Proposed SP-I Scheme (per iteration)} \\ [0.5ex]
 \hline
\end{tabularx}
\begin{tabularx}{\linewidth}{|c|L|L|L|}
    \hline
 Operation & Complex Multiplications & Complex Additions & Total number of operations  \\ [0.5ex]
  \hline
$\boldsymbol{\Omega}^{(n)}_{\mathbf{x}_{p}\hat{\mathbf{x}}_{d}}$
& -& $MN$ & $MN$\\
 \hline
$\mathbf{A}^{(n)}=\big(\boldsymbol{\Omega}^{(n)}_{\mathbf{x}_{p}\hat{\mathbf{x}}_{d}}\big)^H\big(\mathbf{C}^{(n)}_{\boldsymbol{\xi}_{\tilde{\mathbf{w}}}}\big)^{-1} \boldsymbol{\Omega}^{(n)}_{\mathbf{x}_{p}\hat{\mathbf{x}}_{d}}+\mathbf{C}_h^{-1}$
& $Q^2MN+Q^2+Q$& $Q^2(MN-1)+Q^2$ & $2Q^2MN+Q^2+Q$\\
 \hline
$\big(\mathbf{A}^{(n)}\big)^{-1}$ &-&-&$\mathcal{O}(Q^3)$\\
 \hline
 $\mathbf{b}^{(n)}=\big(\boldsymbol{\Omega}^{(n)}_{\mathbf{x}_{p}\hat{\mathbf{x}}_{d}}\big)^H\big(\mathbf{C}^{(n)}_{\boldsymbol{\xi}_{\tilde{\mathbf{w}}}}\big)^{-1}\mathbf{y}$
& $Q(MN+1)$& $Q(MN-1)$ & $2QMN$\\
 \hline
 $\big(\mathbf{A}^{(n)}\big)^{-1}\mathbf{b}^{(n)}$
& $Q^2$& $Q(Q-1)$ & $2Q^2-Q$\\
 \hline
  Data Detection using MP
& -& - & $\mathcal{O}(N_IMNQS)$ \cite{raviteja2018interference}\\
 \hline
\end{tabularx}
\begin{tabularx}{\linewidth}{|L|}
    \hline
 \textbf{Existing EP scheme} \\ [0.5ex]
 \hline
\end{tabularx}
\begin{tabularx}{\linewidth}{|c|L|L|L|}
    \hline
Operation & Complex Multiplications &Complex Additions  & Total number operations  \\ [0.5ex]
 \hline
$|y(k,l)|$
& $(2k_{\text{max}}+1)(l_{\text{max}}+1)$& -& $(2k_{\text{max}}+1)(l_{\text{max}}+1)$\\
 \hline
$\hat{h}_{\text{EP},i}$ for $1\leq i\leq Q$ &$5Q$ &$Q$&$6Q$\\
 \hline
  Data Detection using MP
& -& - & $\mathcal{O}(N_IQS(MN-(2l_{\text{max}}+1)(4k_{\text{max}}+1))\big)$\\
 \hline
\end{tabularx}
\end{table*}
\end{footnotesize}
{We count multiplication/division and addition/subtraction as operations \cite{SinghMJV19}. We see from Table-\ref{tab:a} that the total number of operations required by the proposed SP-NI scheme is $\big(2Q^2+2Q\big)MN+3Q^2+\mathcal{O}(Q^3)+\mathcal{O}(N_IMNQS)$.  We see that the complexity of the proposed SP-I scheme in each iteration varies as $\big(2Q^2+2Q+1\big)MN+3Q^2+\mathcal{O}(Q^3)+\mathcal{O}(N_IMNQS)$. In practice, $Q\ll MN$, we see that the complexity of the proposed SP-NI scheme is $\mathcal{O}(MN)+\mathcal{O}(N_IMNQS)$. The  complexity of  SP-I scheme is $\Big(\mathcal{O}(MN)+\mathcal{O}(N_IMNQS)\Big)N_{\text{SPI}}$, where $N_{\text{SPI}}$ denotes the number of iteration required by the SP-I scheme. We see from Table-\ref{tab:a} that the existing EP added channel estimation scheme requires $(2k_{\text{max}}+1)(l_{\text{max}}+1)+6Q+\mathcal{O}(N_I(MN-\big((2l_{\text{max}}+1)(4k_{\text{max}}+1))QS\big)$, where  $l_{\text{max}}$ and $k_{\text{max}}$ denote the tap corresponding maximum delay and Doppler shift, respectively. Typically, $k_{\text{max}}, l_{\text{max}},Q\ll MN$, the last term in the complexity of the EP scheme dominates. Next section shows numerical comparison of the complexities.}
\vspace{-0.4cm}
    \section{Simulation results}\label{Results}
We now numerically validate the derived analytical results, and the performance of the proposed designs. For this study, we consider an OTFS system with the number of delay bins $M\in\{16,32\}$, the Doppler bins $N\in \left\{16, 32\right\}$, and set the carrier frequency and subcarrier
spacing  as $4$ GHz and $15$ KHz, respectively.  The system uses a rectangular pulse, and BPSK constellation.  We use, similar to \cite{DBLP:conf/globecom/RamachandranC18}, a $5$-tap delay-Doppler channel whose parameters are given in  Table \ref{table:2}. We terminate the i) message passing algorithm, similar to \cite{raviteja2018interference},  using the convergence indicator in  \eqref{eq:conver_I};  ii)  proposed SP-I algorithm when $||\hat{\mathbf{h}}^{(n)}-\hat{\mathbf{h}}^{(n+1)}||^2<10^{-6}$ or when the number of iteration $ 10 $, whichever is achieved earlier. We define the SNR  as $ 1/\sigma_{w}^{2} $.  

{For a fair comparison between the proposed and existing designs, we assume that for each scheme, similar to \cite{TranPTN09}, total power per delay-Doppler frame is same, and we fix it as $MN$. It implies that each delay-Doppler bin $(l,k)$ ($0\leq l \leq M-1$ and $0\leq k \leq N-1$) is assigned a unity power (normalized power).  For the proposed designs, the power per delay-Doppler superimposed symbol in the frame, as shown  in Fig.~\ref{fig:Frame}(b), is $\sigma^2_d+\sigma^2_p=1$, where $\sigma^2_d$ and $\sigma^2_p$ denote data symbol and pilot symbol power respectively. This allows us to define SNR per superimposed symbol as $1/\sigma^2_w$, where $\sigma^2_w$ is the noise variance. The EP design, as shown in Fig.~\ref{fig:Frame}(a), inserts zeros to avoid interference between data and pilot symbols. The total power of $MN$ units in this scheme is therefore distributed as follows.  Its power per data symbol is $\sigma^2_{d,\text{EP}}=\sigma^2_d+\sigma^2_p=1$. This also implies that the SNR per data symbol is $1/\sigma^2_w$, which is equal to that of the proposed SP-aided designs. The pilot power in the EP scheme, however, due to insertion of $(2l_{max}+1)(4k_{max}+1)-1$ zeros around the pilot symbol,  is  $\sigma^2_{p,\text{EP}}=(2l_{max}+1)(4k_{max}+1)$ \cite{DBLP:journals/tvt/RavitejaPH19}. The CPA design employs an entire frame for transmitting pilots, its total pilot power is therefore $MN$. Each of its  delay-Doppler data symbol in the subsequent frame has a power $\sigma^2_{\text{CPA}}=1$.} 
\vspace{-0.5cm}
\begin{table}[htbp]
\centering
\caption{Delay-Doppler channel parameters}
\begin{tabular}{| m{11em} | m{.5cm}| m{1cm}|  m{1cm}| m{1cm}| m{1cm}|} 
 \hline
 Channel tap no. & 1 & 2 & 3 & 4 & 5   \\ 
 \hline
 Delay $(\mu s)$ & 2.08 & 5.20 & 8.328 & 11.46 & 14.80 \\ 
 \hline
 Doppler shift (Hz) & 0 & 470 & 940 & 1410 & 1851  \\
 \hline 
Channel tap power (dB) & 1 & -1.804 & -3.565 & -5.376 & -8.860  \\
  \hline
\end{tabular}
\label{table:2}
\end{table}
\vspace{-0.5cm}
{We notice from Section~\ref{sec:optimal_power} that the optimal pilot and  data powers are a function of SNR, number of channel taps $Q$, channel delay-Doppler profile parameter $\sigma^2_h$, number of delay bins $M$ and the number of Doppler bins $N$.  Table-\ref{table:Optimal} summarizes the optimal pilot power $\sigma^2_{p,\text{opt}}$ and the optimal data power $\sigma^2_{d,\text{opt}}$  for different SNR values obtained using the expression derived in the paragraph below \eqref{eq:SINR_diff}. 
	\begin{table}[htbp]
		\centering
		\caption{{\small Optimal Training and Data Powers with $Q=5$, $M=N=16$ and $\sigma^2_h$ (taken from Table~\ref{table:2}). }}
		\begin{tabular}{| m{5em} | m{1cm}| m{1cm}| m{5em} | m{1cm}| m{1cm}|} 
			\hline
			SNR (dB)& 0 & 5 & 10 & 15 & 20  \\ 
			\hline
			 $\sigma^2_{p,\text{opt}}$ & 0.3020 & 0.3153 & 0.3322 & 0.3479 & 0.3600  \\ 
			\hline
		$\sigma^2_{d,\text{opt}}=1-\sigma^2_{p,\text{opt}}$& 0.6980 & 0.6847 & 0.6678 & 0.6521 & 0.6400 \\		
			\hline
		\end{tabular}
		
		\label{table:Optimal}
	\end{table}
	\vspace{-0.1cm}
	
We see from this table that for the given simulation parameters,  the average optimal pilot and data powers are approximately $0.3$ and $0.7$, respectively. To maximize the SE and to minimize the BER of the proposed SP-aided designs, $30\%$ of the total power should thus be allocated to pilots and  $70\%$ to data, as also numerically verified  in sequel.}
\vspace{-0.6cm}
\subsection{Channel estimator MSE  of the proposed SP-NI and SP-I designs}
\vspace{-0.2cm}
{Figure~\ref{fig:MSE_COMP}(a) shows the BER of the proposed designs as a function of pilot power $\sigma^2_p$. We see that both the schemes yield minimum BER when pilot power $\sigma^2_p=\sigma^2_{p,\text{opt}}=0.3322$. This numerically verifies the results shown in Table-\ref{table:Optimal}. Since $\sigma^2_p+\sigma^2_d =1$, we also observe that the BER of the proposed designs  for i) $\sigma^2_p<\sigma^2_{p,\text{opt}}$, degrades due to the poor channel estimate; ii) for $\sigma^2_p>\sigma^2_{p,\text{opt}}$, increases due to reduced data power.} 

Figure~\ref{fig:MSE_COMP}(b) shows the MSE of the SP-NI channel estimator for different pilot and data power distributions. We see that the increase in the pilot power $\sigma^2_p$ ($\sigma^2_d=1-\sigma^2_p$), which proportionately reduces the data power  $\sigma^2_d$, reduces its MSE. This is because data acts as an interference while estimating channel. 
\vspace{-.6cm}
\begin{figure}[htbp]
	\begin{center}
			\subfloat[]{\includegraphics[scale = 0.48]{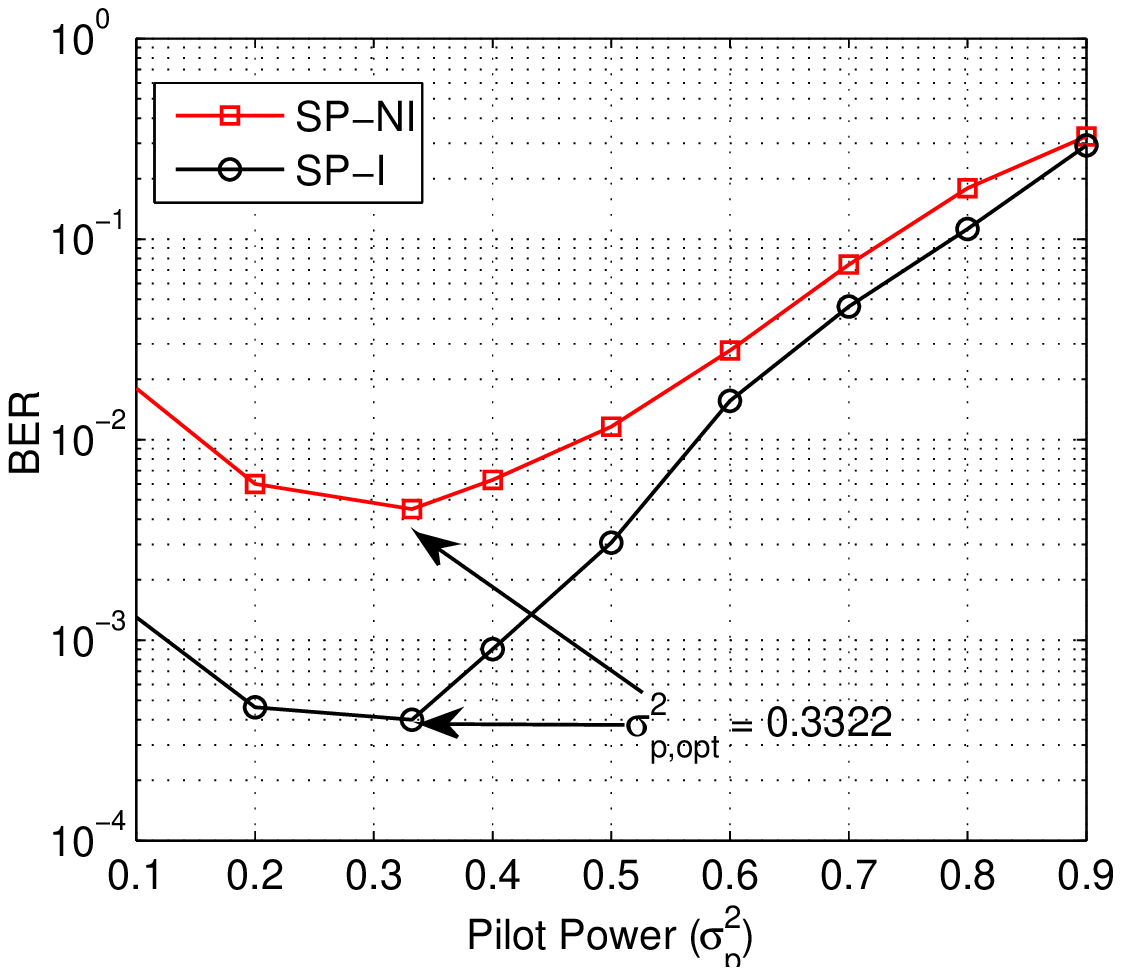}}
		\hfil 
		\subfloat[]{\includegraphics[scale = 0.48]{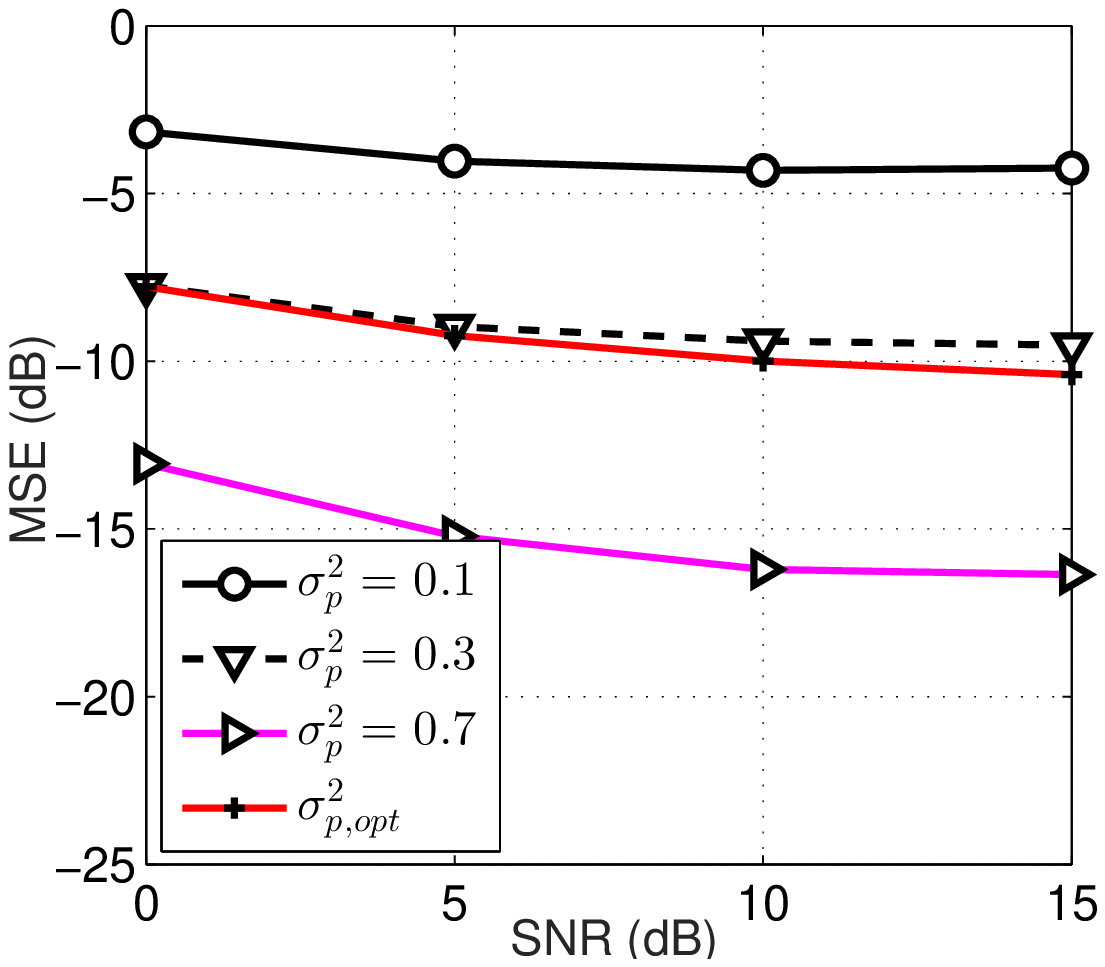}}
		\hfil 
\subfloat[]{\includegraphics[scale = 0.48]{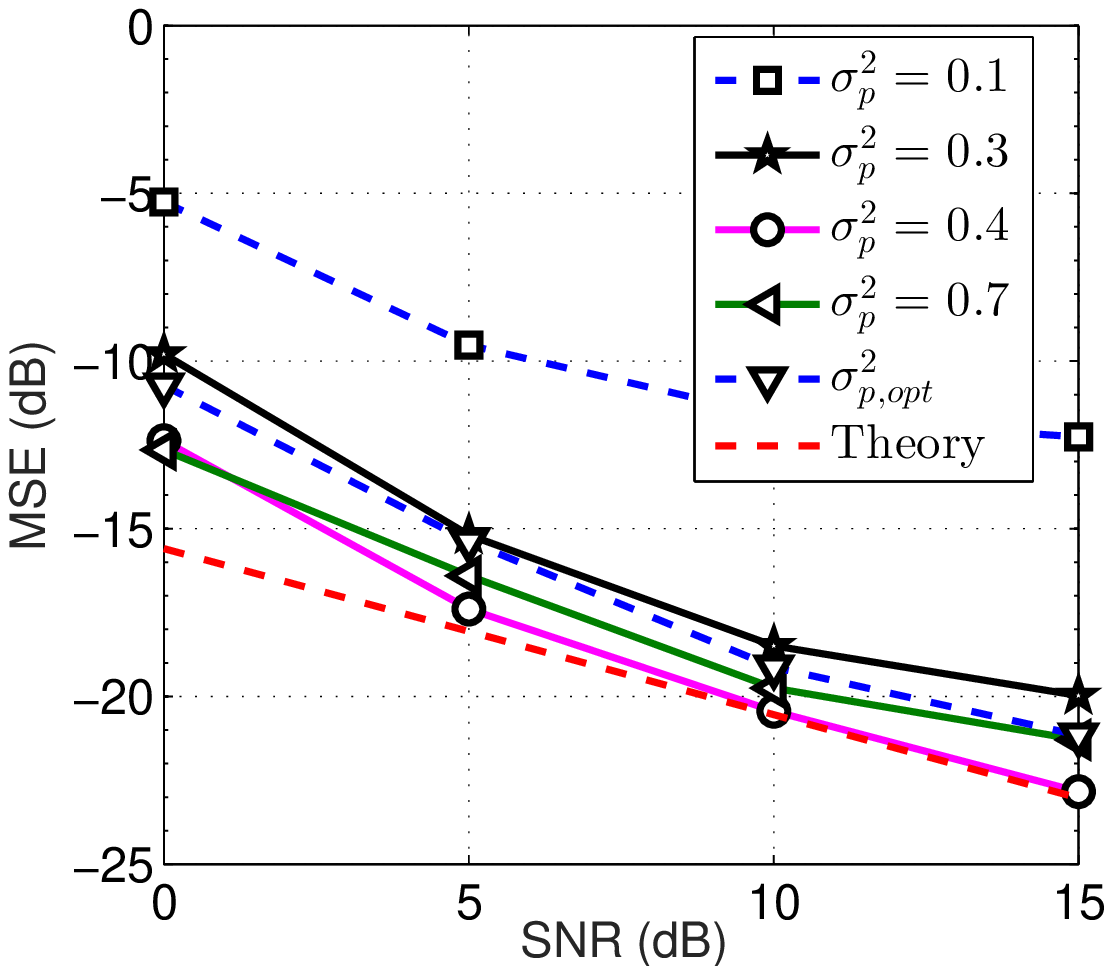}}
	\end{center}	
	\hfil
			\vspace{-.7cm}
	\caption{\small {(a) BER versus pilot power $\sigma^2_p$ for the proposed schemes with $M=N=16$ and SNR $=10$ dB.} Effect of power distribution among pilot and data  on the channel estimation MSE  for  (b)  SP-NI; and (c) SP-I design.}
	\label{fig:MSE_COMP}
	\vspace{-0.7cm}
\end{figure}
The SP-NI design, therefore,  benefits from the increased pilot power and  the  reduced interference power. We see that the optimal pilot power does not minimize the  MSE. This is because it maximizes the data symbols SINR, and not the channel estimator MSE.

Figure~\ref{fig:MSE_COMP}(c) plots the  channel estimator MSE for the SP-I design which iteratively estimates channel and detects data. We also plot its theoretical MSE, which is derived using perfect data knowledge as follows. The received signal in (\ref{eqn:inopy}) can be re-written as 
$\mathbf{y}=\boldsymbol{\Omega}_{\mathbf{x}_{p}\mathbf{x}_{d}} \mathbf{h} + \tilde{\mathbf{w}}$,
with $ \boldsymbol{\Omega}_{\mathbf{x}_{p}\mathbf{x}_{d}}=\boldsymbol{\Omega}_{p} + \boldsymbol{\Omega}_{d} $. The MSE of this estimator is $\tilde{B}_h=\mbox{Tr}\big[\tilde{\boldsymbol{\Sigma}}_{\mathbf{h}}\big]$ \cite{kay1993fundamentals}, where $\tilde{\boldsymbol{\Sigma}}_{\mathbf{h}}=\Big(\boldsymbol{\Omega}^{H}_{\mathbf{x}_{p}\mathbf{x}_{d}} \mathbf{C}_{\tilde{\mathbf{w}}}^{-1} \boldsymbol{\Omega}_{\mathbf{x}_{p}\mathbf{x}_{d}} + \mathbf{C}_{\mathbf{h}}^{-1} \Big)^{-1}$.
The covariance matrices $ \mathbf{C}_{\tilde{\mathbf{w}}} $ and $  \mathbf{C}_{\mathbf{h}} $ are given in (\ref{eqn:Covnoise}) and (\ref{Covh1}), respectively. We see that the optimal pilot power $\sigma^2_{p, \text{opt}}$ does not minimize the MSE. This happens because  $\sigma^2_{p, \text{opt}}$ maximizes the SINR, which does not necessarily minimize the channel estimator MSE. We also see that,  unlike that of the SP-NI design in Fig.~\ref{fig:MSE_COMP}(a), the MSE of the  SP-I design reduces negligibly for $\sigma^2_p> \sigma^2_{p, \text{opt}}$. This is because the SP-I channel estimator MSE depends on the data detection accuracy, which  is a function of the data power~$\sigma^2_d$. We also see that its simulated MSE, with $\sigma^2_p=0.4$ and $\sigma^2_d=0.6$, is close to the theoretical MSE obtained by assuming perfect data availability. This demonstrates the accuracy of the proposed SP-I design.}


\vspace{-0.5cm}
\subsection{BER of the proposed SP-NI and SP-I designs}\label{ber-comp}
\vspace{-0.2cm}
We plot in Fig.~\ref{fig:BER_COMP}(a) and Fig.~\ref{fig:BER_COMP}(b) the BER of the proposed SP-NI and SP-I designs respectively,  for different pilot and data powers. We observe from these plots that the proposed designs, with optimal power allocation, have minimum BER, which also overlaps with BER with $\sigma^2_p=0.3$ and $\sigma^2_d=0.7$. 
	\vspace{-.6cm}
\begin{figure}[htbp]
	\begin{center}
	\subfloat[]{\includegraphics[scale = 0.45]{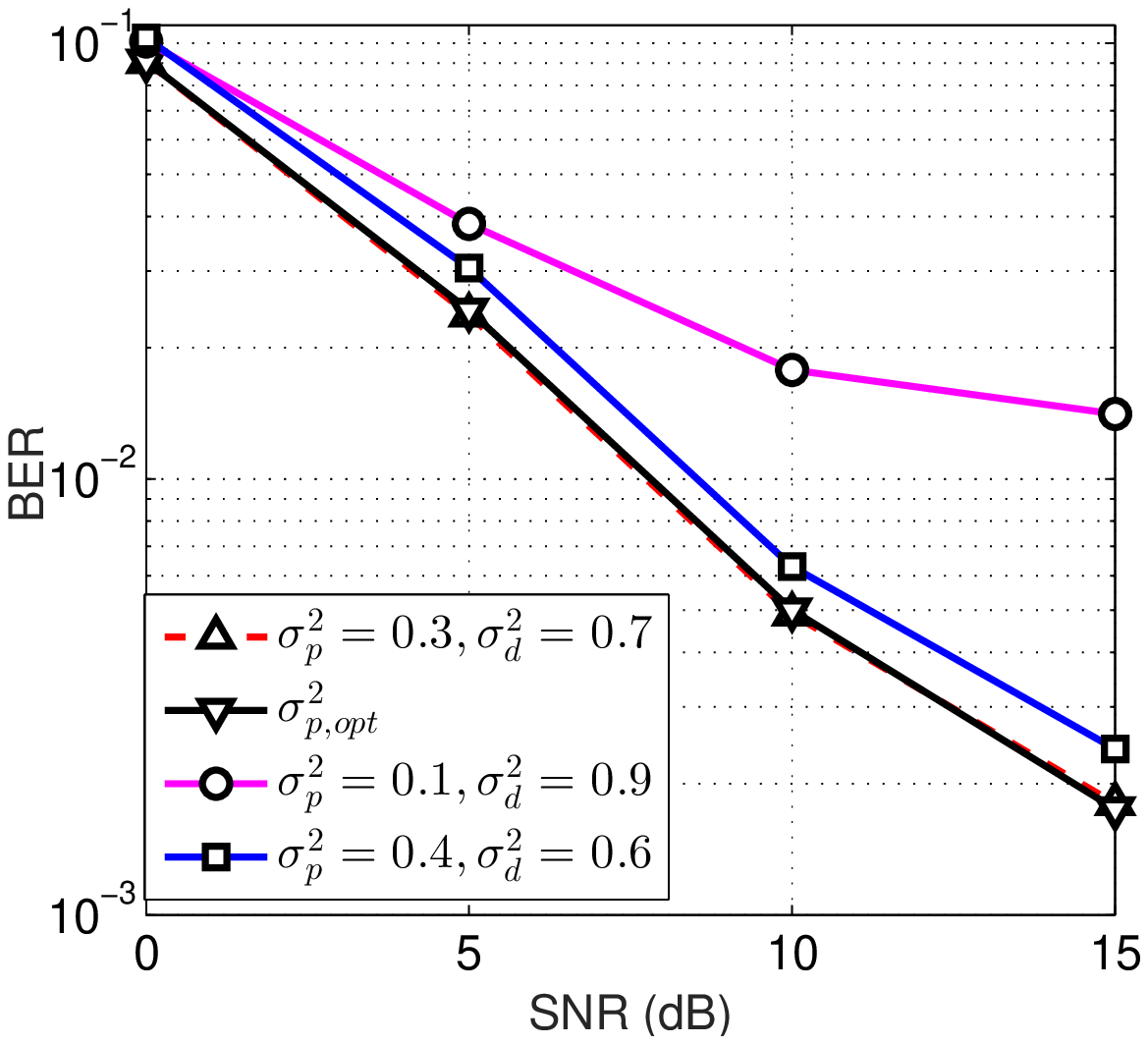}}
	\hfil 
	\hspace{-10pt}\subfloat[]{\includegraphics[scale = 0.47]{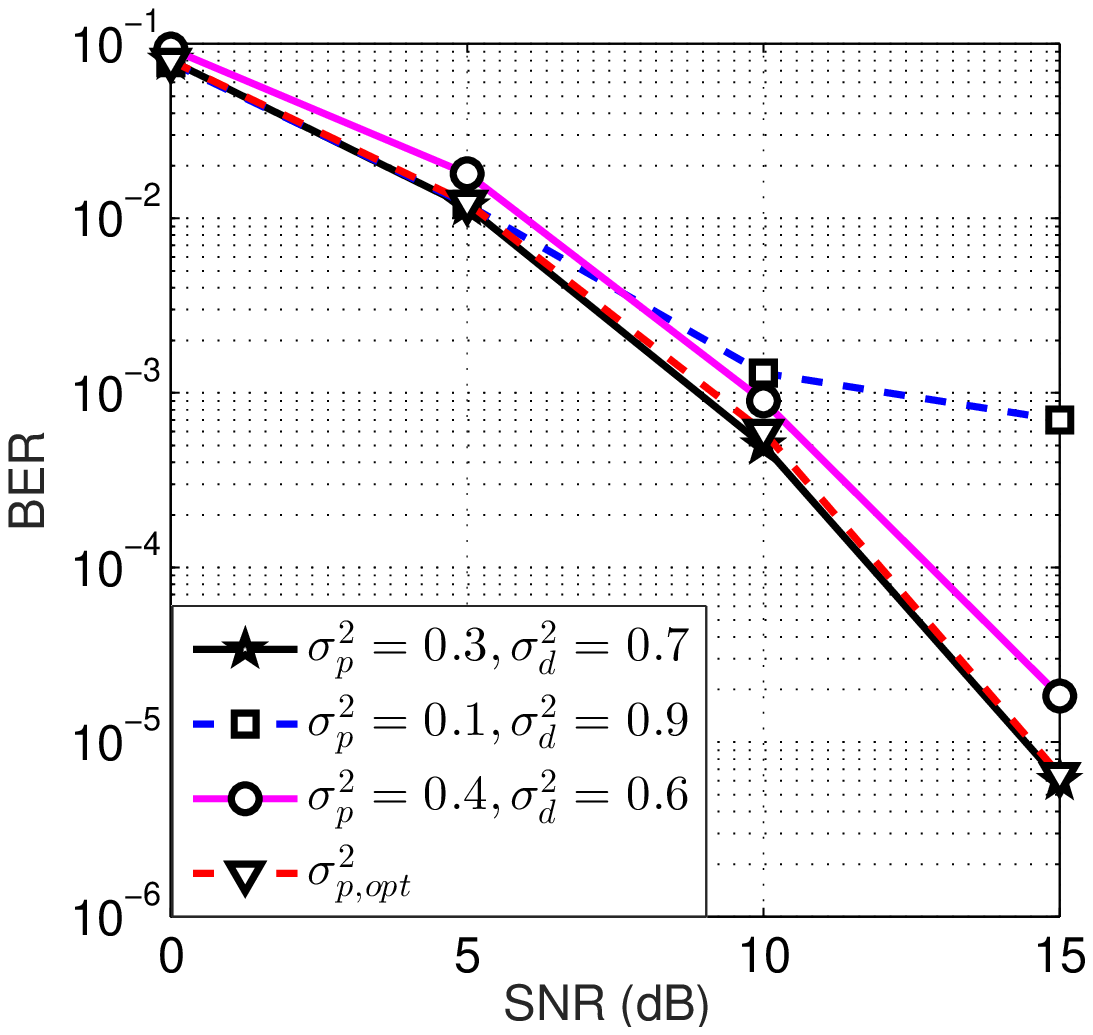}}
	\hfil 								\hspace{-10pt}\subfloat[]{\includegraphics[scale = 0.47]{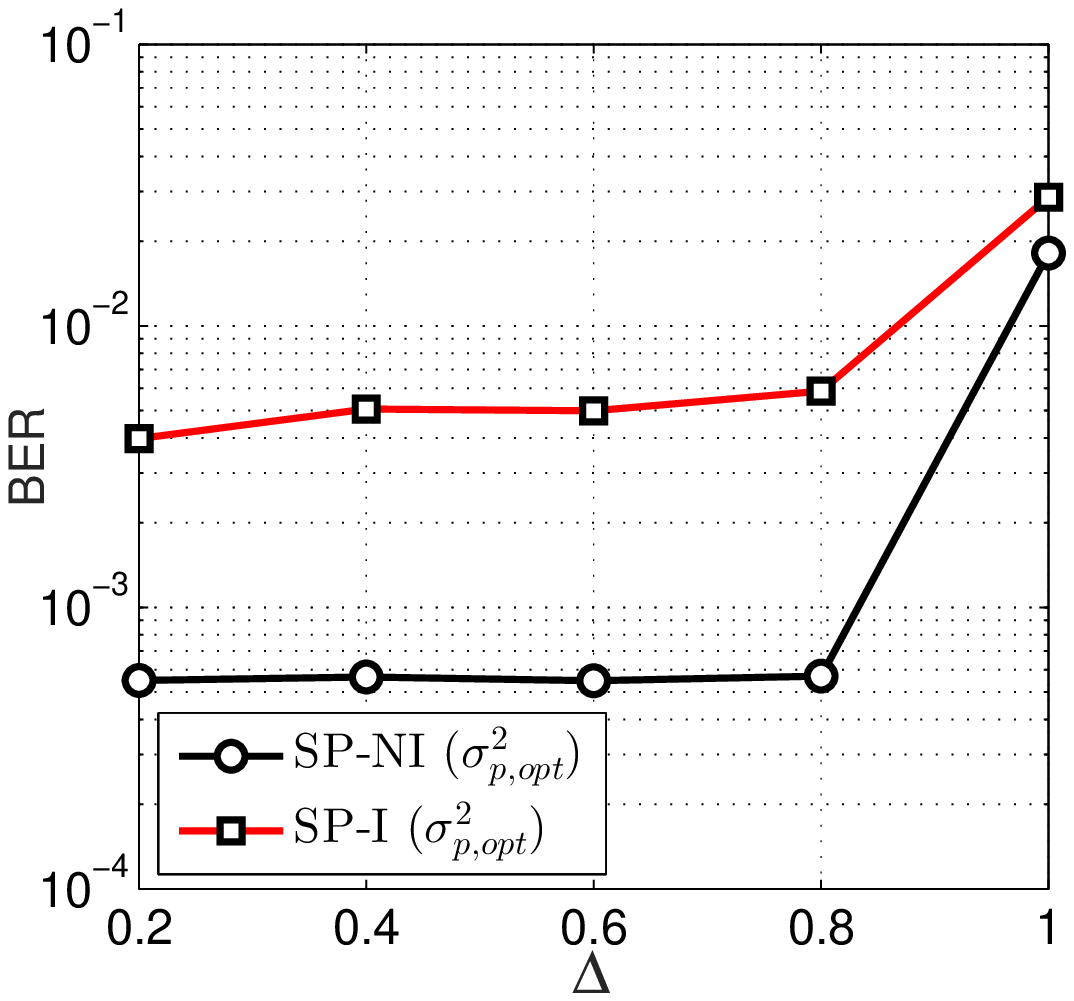}}
	\end{center}	
	\hfil
			\vspace{-.7cm}
	\caption{\small Effect of power distribution among pilots and data on BER of (a)  SP-NI; and (b) SP-I design. (c) {Effect of damping factor $\Delta$ on their BER with SNR $=10$ dB, and  number of message passing iterations $N_I=20$.}}
	\label{fig:BER_COMP}
	\vspace{-0.6cm}
\end{figure}
This validates the optimal pilot power $\sigma^2_{p,\text{opt}}\approx 0.3$ and  data power $\sigma^2_{d,\text{opt}}\approx 0.7$, which is calculated analytically and given in Table-\ref{table:Optimal}. We observe from  Fig.~\ref{fig:BER_COMP}(a) and Fig.~\ref{fig:BER_COMP}(b) that BER of the SP-I design is much lower than the SP-NI design for all SNR values. {The BER gap between the two widens at high SNR values. For example, the SP-I design achieves a BER of $10^{-3}$ at $8.5$ dB, whereas the SP-I design is not able to attain this BER in the operating SNR range.  This is because the the SP-NI design experiences mutual interference between data and pilot symbols, which degrades its performance at high SNR values.}
 
{The message passing algorithm, as shown in \eqref{eq:pmf},  employs damping method for recursively updating pmfs. The damping factor $\Delta\in [0,1]$  controls the convergence of the message passing algorithm \cite{raviteja2018interference,pretti2005message}.  Figure~\ref{fig:BER_COMP}(c) shows the effect of $\Delta$ on the BER of the proposed designs with  the maximum number of iterations for the  convergence of message passing algorithm as $N_I=20$ \cite{raviteja2018interference}. We see that when $0<\Delta\leq 0.8$, BER of both the proposed designs remains constant. This is because  the message passing algorithm converges within $N_I=20$ iterations. For $\Delta>0.8$, their BER  degrade as the algorithm does not  converge within $N_I=20$ iterations.}

\vspace{-0.5cm}
\subsection{SE and BER comparison of the proposed and existing designs}
\vspace{-0.2cm}
This section compares the  SE  of the proposed SP-NI and SP-I  designs with  the state-of-the-art embedded-pilot (EP) design in \cite{DBLP:journals/tvt/RavitejaPH19} and conventional pilot aided (CPA) design in \cite{DBLP:conf/globecom/RamachandranC18}.  The EP design, as shown in Fig.~\ref{fig:Frame}(a), insert zeros between data and pilot, whose number depends on both $l_{max}$ and $k_{max}$ i.e., the tap corresponding to the maximum delay and Doppler shifts, respectively.  The CPA design uses an entire frame for OTFS channel estimation.
The proposed SP-aided designs, as shown in Fig.~\ref{fig:Frame}(b), neither transmit separate pilots nor insert guard symbols (zeros). This radically increase their SE, which we compare next. Before doing that, we calculate the SE  of both EP and CPA designs.
The SE of EP design is  $\mathcal{R}_{\text{EP}} =(1-\eta)\log_2\left(1+\mbox{SINR}_{\text{EP}}\right)$.
The pilot overhead $\eta$ can be calculated using its frame-structure in Fig.~\ref{fig:Frame}(a), and is given as
\begin{equation}\label{eq:ETA}
\eta = \frac{(2l_{max}+1)(4k_{max}+1)}{MN}.
\end{equation}
The SINR of EP scheme, as derived in Appendix~\ref{Appendix_E}, is $\mbox{SINR}_{\text{EP}}=\dfrac{(\sigma^2_h-B_{h,\text{EP}})\sigma^2_{d,\text{EP}}}{\sigma^2_w+\sigma^2_{d,\text{EP}}B_{h,\text{EP}}}$. {The CPA design in \cite{DBLP:conf/globecom/RamachandranC18} uses the first OTFS frame for estimating channel, and the subsequent frame for transmitting data. The value of the pilot overhead parameter $\eta=MN/(2MN)=0.5$.  Its  SE is, therefore, $\mathcal{R}_{\text{CPA}}=\frac{1}{2}\mbox{log}_2(1+\mbox{SINR}_{\text{CPA}})$} with   
$\mbox{SINR}_{\text{CPA}}=\dfrac{(\sigma^2_h-B_{h,\text{CPA}})\sigma^2_{d,\text{CPA}}}{\sigma^2_w+\sigma^2_{d,\text{CPA}}B_{h,\text{CPA}}}$.
Here $B_{h,\text{CPA}}$ is the MSE of  MMSE channel estimator in the CPA design.

The SE of SP-NI scheme is  $\mathcal{R}_{\text{SPNI}}=\mbox{log}_2(1+\mbox{SINR}_{\text{SPNI}})$, where  $\mbox{SINR}_{\text{SPNI}}$ is given in  \eqref{eqn:snr1}.  The SE of SP-I design is  $\mathcal{R}_{\text{SPI}}=\mbox{log}_2(1+\mbox{SINR}_{\text{SPI}})$, where  $\mbox{SINR}_{\text{SPI}}$ is computed using  \eqref{eqn:snr1} as 
 \begin{eqnarray}
 \label{eqn:SPE}
 \mbox{SINR}_{\text{SPI}}= \dfrac{\left(\sigma_{h}^{2}-B_{h}\right)\sigma_{d,\text{opt}}^{2}}{B_{h}\sigma_{d,\text{opt}}^{2}+B_{h}Q\sigma_{p,\text{opt}}^{2}+ \sigma_{w}^{2}}.
 \end{eqnarray}
Here $B_h$ is the MSE of channel estimation in the SP-I design, which is obtained from \eqref{eq:MSE_SPI}, once the Algorithm~\ref{algo:Summary} converges. We calculate the SINR of the proposed designs using optimal powers  $\sigma^2_{p,\text{opt}}$ and $\sigma^2_{d,\text{opt}}$. This is because these powers maximize both SINR and SE of the proposed designs, as verified next. 
Figure~\ref{fig:OH_COMP}(a) shows the effect of power allocation between pilot and data symbols on the SE of the proposed designs. We see that the i) optimal power allocation maximizes their SE; ii)  SP-I design has significantly higher SE that the SP-NI design. This is due to its lower channel estimation MSE}; and iii) {SE of the SP-NI design, unlike the SP-I design,  degrades at high SNR values. This  is due to interference between data and pilots.

Figure~\ref{fig:OH_COMP}(b) compares the SE of the proposed SP-NI and SP-I designs with the EP and CPA designs for $M=N=16$, where $M$ and $N$ are the number of delay and Doppler bins, respectively.  Figure~\ref{fig:OH_COMP}(c) performs the same study for $M=N=32$. We observe from both the figures that the SE of all the designs increases with SNR, which  is not surprising.  We crucially observe that the SE of the proposed SP-I design is significantly higher than the EP design.  {We see that for large values of $l_{max}$ and $k_{max}$, the SE of the EP design degrades so much that even the SP-NI design outperforms it. This is because, as shown in \eqref{eq:ETA}, its pilot overhead increases with $l_{max}$ and $k_{max}$.  The proposed designs, in contrast, avoid this pilot overhead.   We  observe from Fig.~\ref{fig:OH_COMP}(c) that for $l_{max}=4$ and $k_{max}=2$, the SE of the EP design is close to that of the proposed SP-I design, because the pilot overhead $\eta$ decreases with $l_{max}$ and/or $k_{max}$.} 
\vspace{-0.6cm}
\begin{figure}[htbp]
	\begin{center}
		\hspace{-10pt}\subfloat[]{\includegraphics[scale = 0.47]{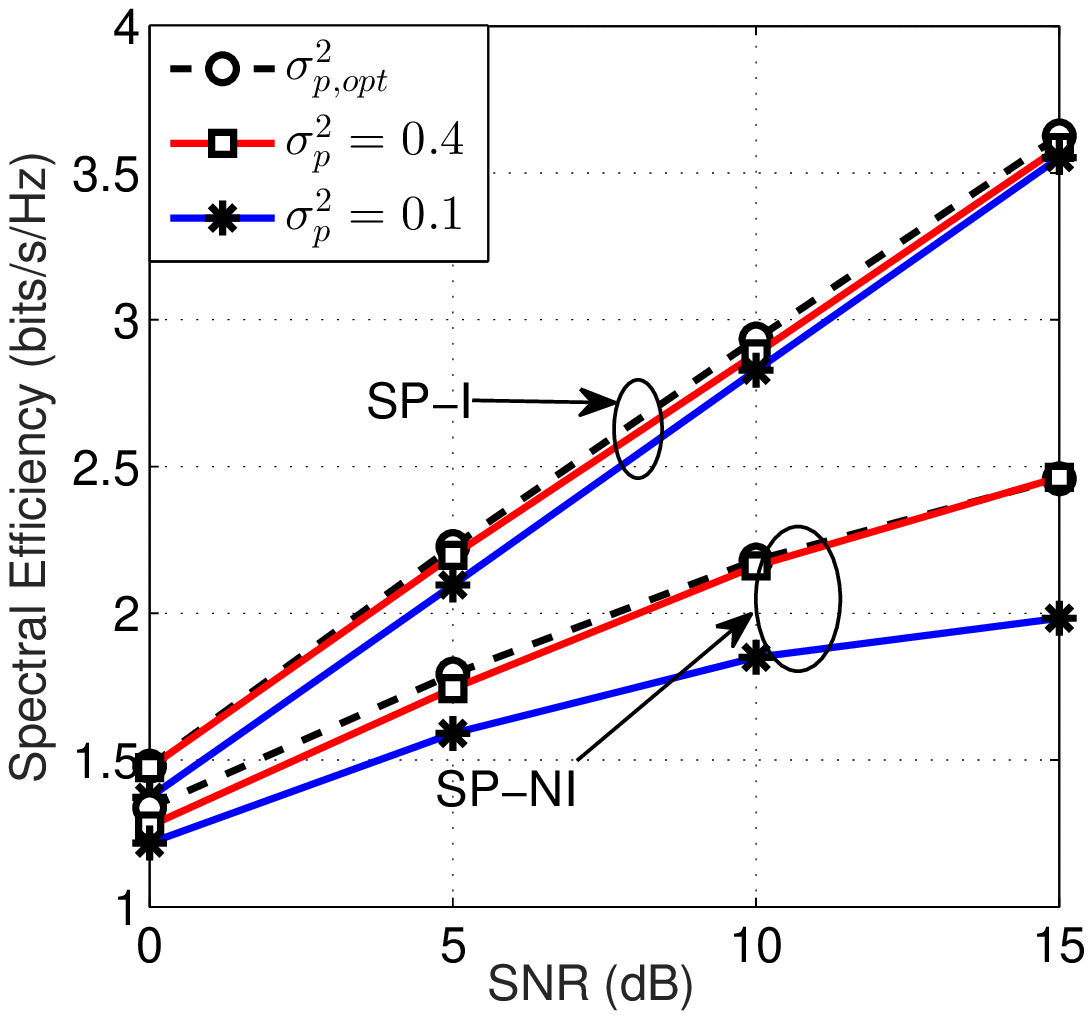}}
		\hfil 
		\hspace{-10pt}\subfloat[]{\includegraphics[scale = 0.47]{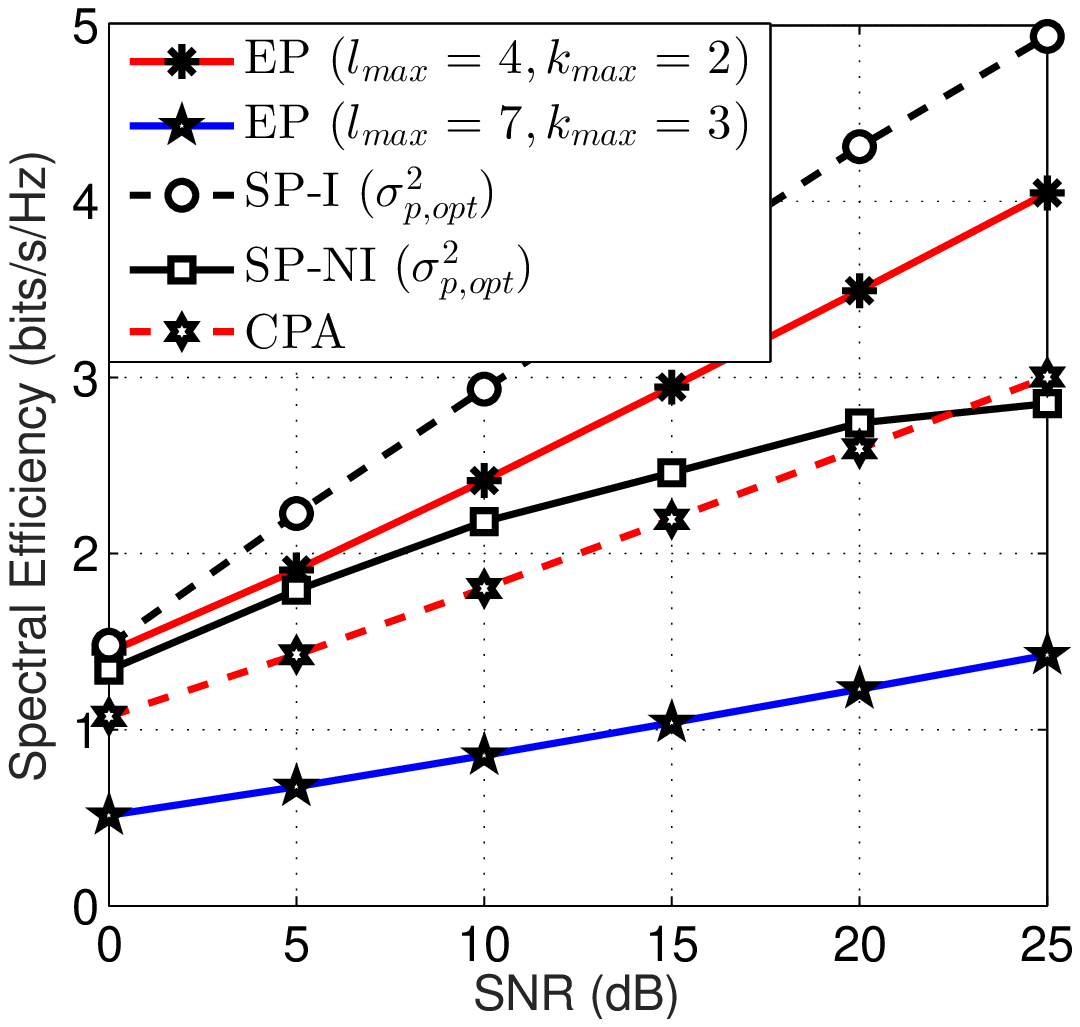}}
		\hfil
		\hspace{-10pt}\subfloat[]{\includegraphics[scale = 0.47]{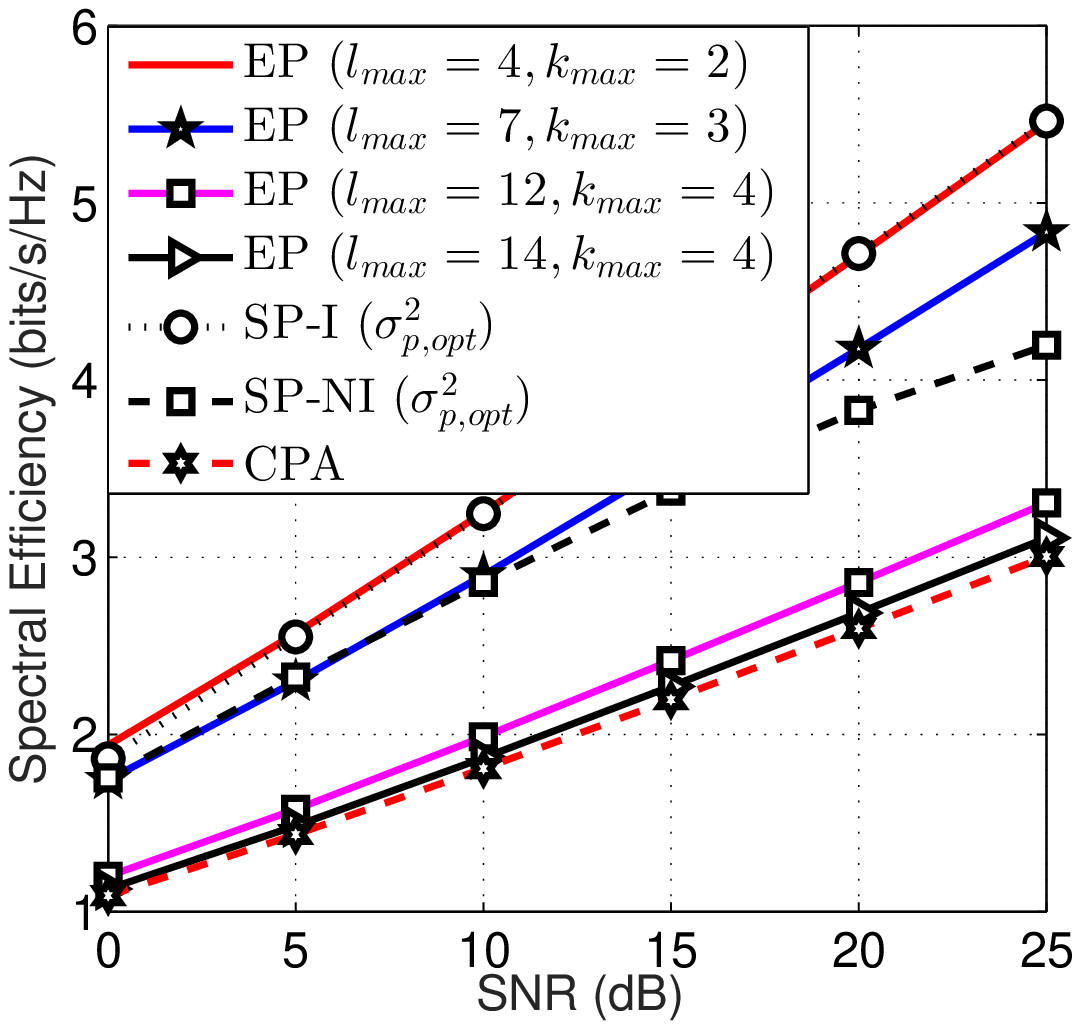}}	
	\end{center}	
	\hfil
	\vspace{-.7cm}
	\caption{\small (a) Effect of  power allocation between data and pilot symbols on the SE of SP-NI and SP-I designs. SE  of the proposed and the existing EP \cite{DBLP:journals/tvt/RavitejaPH19} and CPA \cite{DBLP:conf/globecom/RamachandranC18} designs with (b) $M=N=16$; and (c) $M=N=32$. }
	\label{fig:OH_COMP}
\end{figure}
\vspace{-0.5cm}

 {We also see from Fig.~\ref{fig:OH_COMP}(b) and Fig.~\ref{fig:OH_COMP}(c) that both SP-NI and SP-I designs vastly outperform the CPA design, which  uses one complete frame for transmitting pilots.} The proposed designs, in contrast,  avoid this overhead. The SE of CPA scheme may further degrade in high Doppler scenarios due to channel aging. This is because it  assumes that the channel remains constant for at least two consecutive frames.

Figure~\ref{fig:SE_COMP}(a) shows the SE of the proposed SP-NI and SP-I designs, and the EP design,  by varying the number of delay and Doppler bins $M$ and $N$ respectively as $M=N$, for different values of $l_{max}$ and $k_{max}$.  We observe that the SE of both the proposed designs increases with  $M$ and $N$. This is because, {as shown in \eqref{eqn:mse31}, the channel estimation MSE reduces with increasing $M$ and $N$. The improved channel estimates, which initialize the proposed SP-I design, boosts it SINR and consequently the SE.}  The pilot overhead $\eta$ for the EP design reduces with increase in $M=N$, and its the SE consequently approaches that of the SP-I design.
\vspace{-0.7cm}
\begin{figure}[h]
	\begin{center}
		\subfloat[]{\includegraphics[scale = 0.45]{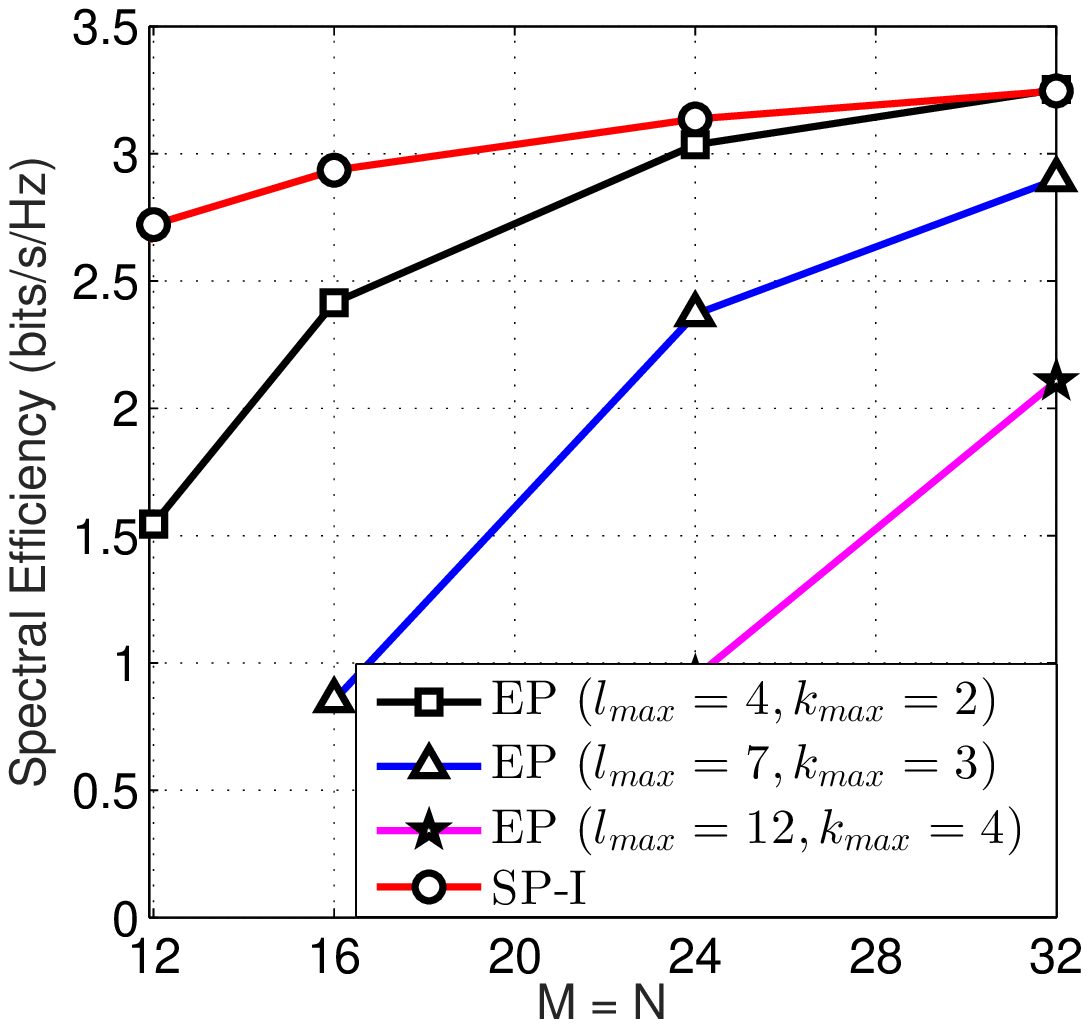}}
		\hfil 
		\subfloat[]{\includegraphics[scale = 0.45]{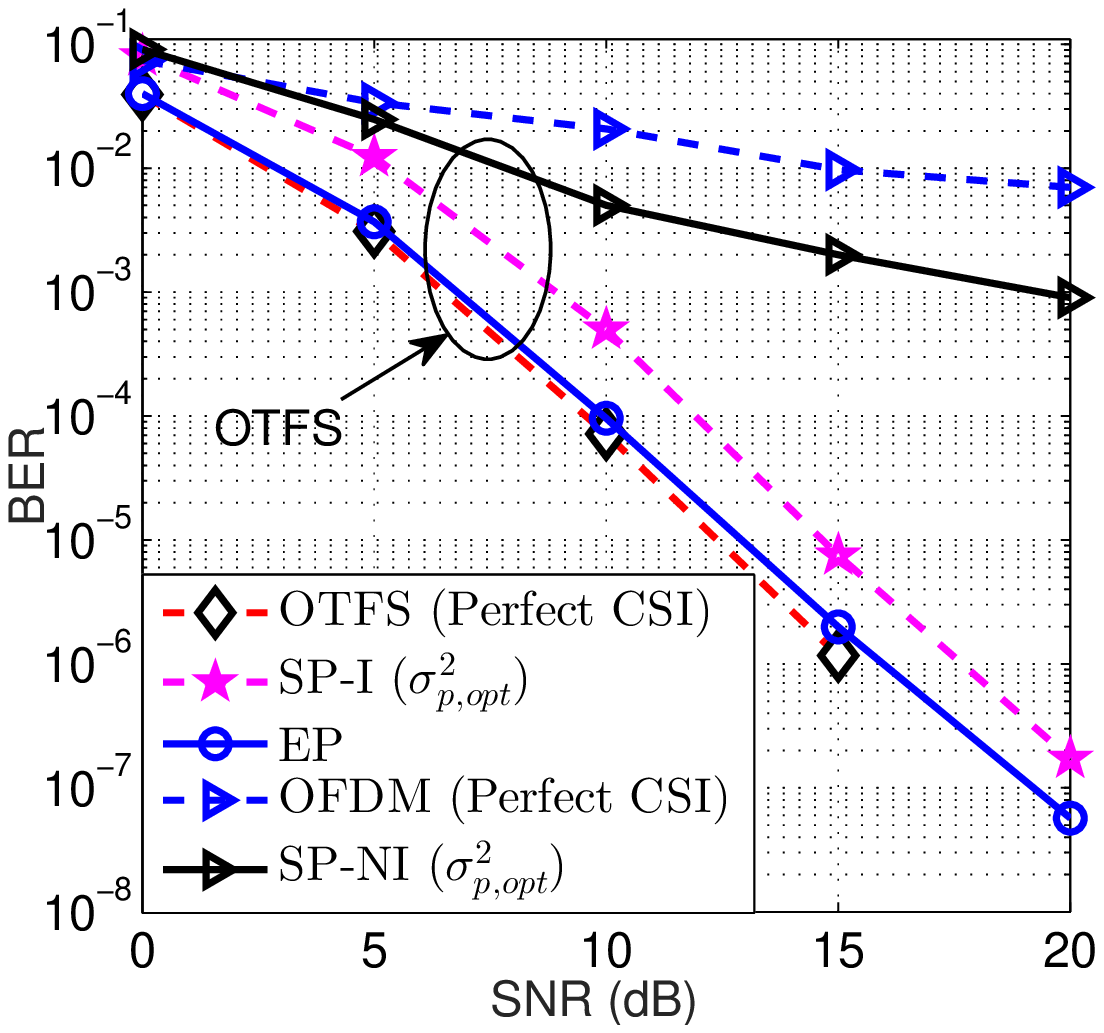}}
		\hfil 
	\subfloat[]{\includegraphics[scale = 0.49]{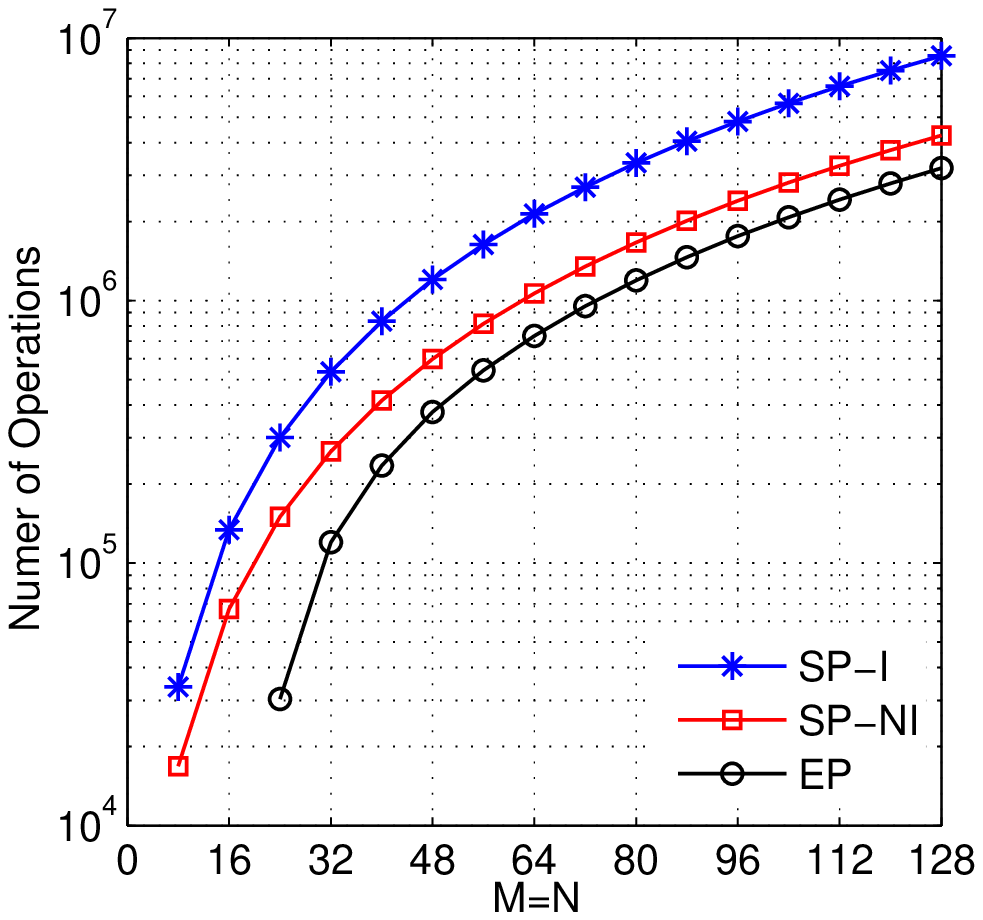}}
	\end{center}	
	\hfil
	\vspace{-.6cm}
	\caption{\small (a)  SE versus $M=N$ comparison  of the proposed SP-I and the EP design in \cite{DBLP:journals/tvt/RavitejaPH19} at SNR = $10$ dB. (b) {BER comparison of the proposed and existing designs with $M=N=16$}. (c) {Computational complexity of the proposed and existing design with $Q=5$, $S=2$, $N_I=20$, $N_{\text{SPI}}=2$, $k_{\text{max}}=12$ and $l_{\text{max}}=4$.}}
	\label{fig:SE_COMP}
	\vspace{-0.7cm}
\end{figure}

Figure~\ref{fig:SE_COMP}(b) compares the BER of the proposed SP-NI and SP-I designs with the EP design. To benchmark the performance, we also plot the BER of the OTFS and  OFDM systems with perfect receive CSI.  We observe that the BER of the OTFS system, with the proposed SP-based designs, is significantly lower than that of the OFDM system. {We also see that EP-based design in \cite{DBLP:journals/tvt/RavitejaPH19} is $\approx 1.5$ dB  superior to the proposed SP-I design in terms of BER. This is  because, unlike the EP scheme, the power per symbol in the proposed designs is divided between pilot and data symbols. The power allocated to a data symbol in the proposed designs,  as shown in optimal power allocation in Table~\ref{table:Optimal}, is $\approx 70\%$. This is $\approx 30\%$ lower than the power allocated to data symbols in the EP scheme. \textit{We, however, crucially note that the proposed designs, as shown earlier, has significantly higher spectral efficiency than the EP-based scheme in \cite{DBLP:journals/tvt/RavitejaPH19} and the conventional pilot-aided (CPA) scheme in \cite{DBLP:conf/globecom/RamachandranC18}.}}

{Figure~\ref{fig:SE_COMP}(c)  numerically compares the complexity of the proposed designs with the existing EP scheme. We see that complexity of the i) SP-NI scheme is marginally higher than the EP scheme;  ii) SP-I scheme is slightly higher than both SP-NI and  EP schemes. This is because the proposed SP-I scheme also exploits data symbols to improve channel estimation accuracy. The SE of the proposed SP-NI and SP-I designs, however as shown in Fig.~\ref{fig:OH_COMP}(b) and Fig.~\ref{fig:OH_COMP}(c), is significantly higher than the EP scheme.}

\vspace{-0.4cm}
\section{Conclusions}\label{Conclusion} 
We proposed superimposed pilot (SP)-aided non-iterative (SP-NI) and iterative (SP-I) designs for estimating channel and detecting data. These designs superimpose pilots on to data symbols and, contrary to the existing OTFS channel estimation and data detection designs, do not incur the consequent SE loss. Another key advantage of the proposed designs is that they exploit the OTFS channel sparsity in the delay-Doppler domain by using computationally-efficient message passing algorithm for detecting data. For the proposed designs, we derived a lower bound on the SINR, which is used to  optimally allocate power between data and pilot symbols. We showed that this optimal power allocation minimizes the BER and  maximizes the SE. We also demonstrated that the proposed SP-I design, with a negligible BER degradation, has a significantly higher SE than the existing state-of-the-art designs. This work did not consider data-dependent SP schemes and mean-removal based techniques. Future lines of this work may consider these ideas.
\appendices
\vspace{-0.5cm}
\section{}\label{Lamma_proof}
Using (\ref{eqn:data}), the covariance matrix of $ \boldsymbol{\Omega}_{d} $  can be evaluated~as 
 \begin{eqnarray}
 \label{eqn:covardata}
 \mathbb{E}\left\{\boldsymbol{\Omega}_{d}\boldsymbol{\Omega}_{d}^{H} \right\}= \sum_{i=1}^{Q}\boldsymbol{\Gamma}_{i}\mathbb{E}\left\{\mathbf{x}_{d}\mathbf{x}_{d}^{H}\right\}\boldsymbol{\Gamma}_{i}^{H}
\label{eqn:Omega3}= \sigma_{d}^{2}\sum_{i=1}^{Q} \boldsymbol{\Gamma}_{i}\boldsymbol{\Gamma}_{i}^{H},
 \end{eqnarray}
where we have used: $ \mathbb{E}\left\{\mathbf{x}_{d}\mathbf{x}_{d}^{H} \right\}=\sigma_{d}^{2}\mathbf{I}_{MN} $. The  $ \boldsymbol{\Gamma}_{i}\boldsymbol{\Gamma}_{i}^{H} $ in \eqref{eqn:Omega3} can be evaluated as 
\begin{eqnarray}
\label{eqn:Gamavalue}
\boldsymbol{\Gamma}_{i}\boldsymbol{\Gamma}_{i}^{H}=\mathbf{B}_{\text{rx}}\boldsymbol{\Theta}_{i}\mathbf{B}_{\text{tx}} \mathbf{B}_{\text{tx}}^{H}\boldsymbol{\Theta}_{i}^{H}\mathbf{B}_{\text{rx}}^{H}  = \mathbf{B}_{\text{rx}}\boldsymbol{\Pi}^{l_{i}}\left(\boldsymbol{\Pi}^{l_{i}}\right)^{H}\mathbf{B}_{\text{rx}}^{H}.
\end{eqnarray}
By singular valued decomposition \cite{golub1971singular}, it can be easily shown that the cyclic-shift matrix $ \boldsymbol{\Pi} $ satisfies the following property
$\boldsymbol{\Pi}^{l_{i}}\left(\boldsymbol{\Pi}^{l_{i}} \right)^{H}=\mathbf{I}_{MN}$.
Using this property, we get
$\boldsymbol{\Gamma}_{i}\boldsymbol{\Gamma}_{i}^{H}= \mathbf{B}_{\text{rx}}\mathbf{B}_{\text{rx}}^{H} = \mathbf{I}_{MN}.$
Substituting the last equation in (\ref{eqn:covardata}), we get
\begin{eqnarray}\label{eqn:omegadap}
\mathbb{E}\left\{\boldsymbol{\Omega}_{d}\boldsymbol{\Omega}_{d}^{H} \right\}= \sigma_{d}^{2}\sum_{i=1}^{Q} \mathbf{I}_{MN}=\sigma_{d}^{2}Q\mathbf{I}_{MN}.
\end{eqnarray}
\vspace{-0.9cm}
\section{}\label{Lamma_proof_2}
Since $ \mathbf{x}_{d} $ is statistically independent of $ \mathbf{w} $, the covariance matrix of $ \tilde{\mathbf{w}}_d $ can be evaluated as
\begin{eqnarray}
\label{eqn:Covnoise2}
\mathbb{E}\left\{\tilde{\mathbf{w}}_{d}\tilde{\mathbf{w}}_{d}^{H} \right\}=\mathbb{E}\left\{\boldsymbol{\Omega}_{d}\mathbf{h}\mathbf{h}^{H}\boldsymbol{\Omega}_{d}^{H} \right\}+\mathbb{E}\left\{\tilde{\mathbf{w}}\tilde{\mathbf{w}}^{H} \right\}. 
\end{eqnarray}
The first part of the above expression can be evaluated as 
\begin{align}\label{eqn:expectation2}
\mathbb{E}\left\{\boldsymbol{\Omega}_{d}\mathbf{h}\mathbf{h}^{H}\boldsymbol{\Omega}_{d}^{H} \right\} = \mathbb{E}_{\mathbf{x}_{d}|\mathbf{h}}\left\{ \boldsymbol{\Omega}_{d}\mathbb{E}_{\mathbf{h}}\left\{\mathbf{h}\mathbf{h}^{H} \right\} \boldsymbol{\Omega}^{H}_{d}\right\}  \stackrel{(a)}{=} \left(\sum_{i=1}^{Q}\sigma_{h_{i}}^{2}\right)\sigma_{d}^{2}\mathbf{I}_{MN}.
\end{align}
Equality $ \left(a\right) $  follows from Lemma~\ref{Lemma21} and \eqref{eqn:omegadap}, respectively. Substituting (\ref{eqn:expectation2}) and (\ref{eqn:Covnoise}) in (\ref{eqn:Covnoise2}), we get
$\mathbb{E}\left\{\tilde{\mathbf{w}}_{d}\tilde{\mathbf{w}}_{d}^{H} \right\}=\left( \left(\sum_{i=1}^{Q}\sigma_{h_{i}}^{2}\right)\sigma_{d}^{2} + \sigma_{w}^{2}  \right)\mathbf{I}_{MN}.
$
\vspace{-0cm}
\section{}\label{Appendix_C}
The covariance matrix of the vector $ \boldsymbol{\xi}_{\tilde{\mathbf{w}}}^{\left(n\right)} $ can be evaluated using (\ref{eqn:xi1}) as 
\begin{eqnarray}
\label{eqn:Covxiw45}
\mathbf{C}_{\boldsymbol{\xi}_{\tilde{\mathbf{w}}}}^{\left(n\right)}
\stackrel{(b)}{=}& \mathbb{E}\left\{\boldsymbol{\Xi }^{\left(n\right)}_{\mathbf{x}_{d}}\mathbf{h}\mathbf{h}^{H}\left(\boldsymbol{\Xi }^{\left(n\right)}_{\mathbf{x}_{d}}\right)^{H} \right\}+ \mathbb{E}\left\{\tilde{\mathbf{w}}\tilde{\mathbf{w}}^{H} \right\},
\end{eqnarray}
where the equality $ \left(b\right) $ holds since the data symbols, channel parameters and noise samples are statistically independent. Evaluating $ \mathbb{E}\left\{\boldsymbol{\Xi }^{\left(n\right)}_{\mathbf{x}_{d}}\mathbf{h}\mathbf{h}^{H}\left(\boldsymbol{\Xi }^{\left(n\right)}_{\mathbf{x}_{d}}\right)^{H} \right\} $ using Lemma-\ref{Lemma1}, we get
\begin{align}\label{eqn:ExpXi12}
 \mathbb{E}\left\{\boldsymbol{\Xi }^{\left(n\right)}_{\mathbf{x}_{d}}\mathbf{h}\mathbf{h}^{H}\left(\boldsymbol{\Xi }^{\left(n\right)}_{\mathbf{x}_{d}}\right)^{H} \right\}=\mathbb{E}\left\{\boldsymbol{\Omega}_{d}\mathbf{C}_{\mathbf{h}}\boldsymbol{\Omega}^{H}_{d} \right\}+\mathbb{E}\Big\{\widehat{\boldsymbol{\Omega}}^{\left(n\right)}_{d}\mathbf{C}_{\mathbf{h}}\left(\widehat{\boldsymbol{\Omega}}^{\left(n\right)}_{d}\right)^{H}\Big\}.
\end{align}
Substituting \eqref{eqn:expectation2} in \eqref{eqn:ExpXi12}, we get $ \mathbb{E}\left\{\boldsymbol{\Xi }^{\left(n\right)}_{\mathbf{x}_{d}}\mathbf{h}\mathbf{h}^{H}\left(\boldsymbol{\Xi }^{\left(n\right)}_{\mathbf{x}_{d}}\right)^{H} \right\} =2\left(\sum_{i=1}^{Q}\sigma^{2}_{h_{i}} \right)\sigma_{d}^{2}\mathbf{I}_{MN} $.
Finally, substitution of this result  and (\ref{eqn:Covnoise}) in (\ref{eqn:Covxiw45}) yields the desired result in (\ref{eqn:Covxiw1}). 
\vspace{-0.3cm}
\section{}\label{Appendix_D}
{We have, by using \eqref{eqn:pilot}
\begin{align} \label{eqn:tracepilot1}
\mbox{Tr}\left(\boldsymbol{\Omega}_{p}\boldsymbol{\Omega}_{p}^{H}\right)&=\sum_{i=1}^{Q}\mbox{Tr}\left(\boldsymbol{\Gamma}_{i}\mathbf{x}_{p}\mathbf{x}_{p}^{H}\boldsymbol{\Gamma}_{i}^{H}\right)=\mbox{Tr}\left(\boldsymbol{\Omega}_{p}\boldsymbol{\Omega}_{p}^{H}\right)
=\sum_{i=1}^{Q} \mbox{Tr}\left(\boldsymbol{\Gamma}_{i}^{H}\boldsymbol{\Gamma}_{i}\mathbf{x}_{p}\mathbf{x}_{p}^{H}\right)= QMN\sigma_{p}^{2}. 
\end{align}
{We next use the following result from \cite{ulukok2010some}: For any positive definite matrix $ \mathbf{A} \in \mathbb{C}^{m \times m} $, we have 
$	\mbox{Tr}\left(\mathbf{A}^{-1}\right) \geq \dfrac{m^{2}}{\mbox{Tr}\left(\mathbf{A}\right)}$.
Using this result and \eqref{eqn:incov},
$B_{h, \: \text{NI}}=\mbox{Tr}\left(\boldsymbol{\Sigma}_{\text{NI}}\right) \geq \frac{Q^{2}}{\mbox{Tr}\left(\boldsymbol{\Omega}_{p}^{H}\mathbf{C}^{-1}_{\tilde{\mathbf{w}}_d} \boldsymbol{\Omega}_{p}+ \mathbf{C}^{-1}_{\mathbf{h}}\right)}$.
Its  denominator is simplified next.
\begin{align}
\label{eqn:mse2}
\mbox{Tr}\left(\boldsymbol{\Omega}_{p}^{H}\mathbf{C}^{-1}_{\tilde{\mathbf{w}}_d} \boldsymbol{\Omega}_{p}+ \mathbf{C}^{-1}_{\mathbf{h}}\right)  
&\stackrel{(a)}{=} \dfrac{\mbox{Tr}\left(\boldsymbol{\Omega}_{p}^{H}\boldsymbol{\Omega}_{p}\right)}{\sigma_{h}^{2}\sigma_{d}^{2}+\sigma_{w}^{2}}+ \sum_{i=1}^{Q}\dfrac{1}{\sigma_{h_{i}}^{2}} = \dfrac{QMN\sigma_{p}^{2}}{\sigma_{h}^{2}\sigma_{d}^{2}+\sigma_{w}^{2}}+ \tilde{\sigma}_{h}^{2}.
\end{align}
Here $ (a) $ follows from \eqref{eqn:covariance1} and Lemma \ref{Lemma21}, and $ \tilde{\sigma}_{h}^{2}=\sum_{i=1}^{Q}\left(1/\sigma_{h_{i}}^{2}\right) $. We get the desired result in Lemma~\ref{Lemma4} by substituting \eqref{eqn:mse2} in $B_{h, \: \text{NI}}$. 
\section{}\label{Appendix_E}
Equation (2) of \cite{DBLP:journals/tvt/RavitejaPH19}, similar to \eqref{eq:y_kl}, can be written as 
\begin{align}
y\left[l,k\right]&=\sum_{i=1}^{Q}\hat{h}_{\text{EP},i}\alpha_{i}\left(l,k\right) x_{d, \text{EP}}\left[\left(l-l_{i}\right)_{M}, \left(k-k_{i}\right)_{N}\right]+ \sum_{i=1}^{Q}\left(h_{i}-\hat{h}_{\text{EP},i}\right)\alpha_{i}\left(l,k\right)\nonumber \\
&\times x_{d,\text{EP}}\left[\left(l-l_{i}\right)_{M}, \left(k-k_{i}\right)_{N}\right] +w\left[l,k\right] 
= \tilde{\mathbf{x}}_{d, \text{EP}}^{T}\hat{\mathbf{h}}_{\text{EP}}^{\alpha}+\tilde{v}\left[l,k\right].
\end{align}
Here $\hat{h}_{\text{EP},i}\alpha_{i}$ is the MMSE channel estimate corresponding to the $i$th delay-Doppler path, which is obtained using the EP scheme proposed in \cite{DBLP:journals/tvt/RavitejaPH19}. The noise plus interference term $ \tilde{v}\left[l,k\right]=\tilde{\mathbf{x}}_{d}^{T}\tilde{\mathbf{h}}_{\text{EP}}^{\alpha}+w\left[l,k\right]  $. Here, the $ i $th element of $ \hat{\mathbf{h}}_{\text{EP}}^{\alpha}\in \mathbb{C}^{Q\times 1} $ and  $ \tilde{\mathbf{h}}_{\text{EP}}^{\alpha}\in \mathbb{C}^{Q\times 1} $, are $ \hat{h}_{\text{EP},i}\alpha_{i}\left(l,k\right) $ and $ \left(h_{i}-\hat{h}_{\text{EP},i}\right)\alpha_{i}\left(l,k\right) $, respectively. The SINR of the $(l,k)$th received symbol in the delay-Doppler domain can be formulated from the above expression as
\begin{align}
\label{eq:SINREP}
\text{SINR}_{\text{EP},l,k}=\dfrac{\mathbb{E}\left\{\left|\tilde{\mathbf{x}}_{d,\text{EP}}^{T}\hat{\mathbf{h}}_{\text{EP}}^{\alpha} \right|^{2}\right\}}{\mathbb{E}\left\{\left|\tilde{v}\left[l,k\right] \right|^{2}\right\}}.
\end{align}
We first simplify the numerator of this SINR as 
\begin{align}
\label{eq:EPsignpow}
\mathbb{E}\left\{\left|\tilde{\mathbf{x}}_{d,\text{EP}}^{T}\hat{\mathbf{h}}_{\text{EP}}^{\alpha} \right|^{2}\right\}=\mathbb{E}\left\{\left(\hat{\mathbf{h}}_{\text{EP}}^{\alpha}\right)^{H}\mathbb{E}\left\{\tilde{\mathbf{x}}_{d,\text{EP}}^{*}\tilde{\mathbf{x}}_{d,\text{EP}}^{T}\right\}\hat{\mathbf{h}}_{\text{EP}}^{\alpha}\right\}\stackrel{\left(a\right)}{=}\sigma_{d,\text{EP}}^{2}\mathbb{E}\left\{\left\|\hat{\mathbf{h}}_{\text{EP}} \right\|^{2}\right\}.
\end{align}
Equality $ (a) $ is because $ \mathbb{E}\left\{\tilde{\mathbf{x}}_{d,\text{EP}}^{*}\tilde{\mathbf{x}}_{d,\text{EP}}^{T}\right\}=\sigma_{d,\text{EP}}^{2}\mathbf{I}_{Q} $.The expression $ \mathbb{E}\big\{\|\hat{\mathbf{h}}_{\text{EP}}\|^{2}\big\} $ is next evaluated using  \eqref{eqn:trhhathhatHe}  as $\mathbb{E}\big\{\|\hat{\mathbf{h}}_{\text{EP}}\|^{2}\big\} =\sigma_{h}^{2}-B_{h,\text{EP}}$, where $ B_{h,\text{EP}} $ denotes the MSE of the MMSE channel estimator in EP-based design in \cite{DBLP:journals/tvt/RavitejaPH19}, which is calculated as \cite{kay1993fundamentals}: 
$B_{h,\text{EP}}= \sum_{i=1}^{Q}\dfrac{\sigma^2_w\sigma^2_{h_{i}}}{\sigma^2_{p,\text{EP}}\sigma^2_{h_{i}}+\sigma^2_w}$.
The numerator of $\text{SINR}_{\text{EP},l,k}$ expression is simplified using the above equation as $
\mathbb{E}\big\{|\tilde{\mathbf{x}}_{d,\text{EP}}^{T}\hat{\mathbf{h}}_{\text{EP}}^{\alpha} |^{2}\big\}= \sigma_{d,\text{EP}}^{2}\left(\sigma_{h}^{2}-B_{h,\text{EP}}\right)$. The denominator of SINR in \eqref{eq:SINREP} is given as $ \mathbb{E}\left\{\left|\tilde{v}\left[l,k\right] \right|^{2}\right\}
 =\sigma_{d,\text{EP}}^{2}B_{h,\text{EP}}+\sigma_{w}^{2}.$ Substitution of these terms in \eqref{eq:SINREP} yields the desired SINR expression in \eqref{eq:ETA}.

\ifCLASSOPTIONcaptionsoff
  \newpage
\fi


\bibliographystyle{IEEEtran}
\vspace{-0.5cm}
\bibliography{IEEEabrv,OTFS_SI}

\end{document}